\documentclass[aps,showpacs,nofootinbib,superscriptaddressn,twocolumn,amssymb,amsmath,amsfonts]{revtex4}
\usepackage{latexsym}
\usepackage[latin1]{inputenc}
\usepackage[dvips]{graphicx}
\usepackage{color}

\newcommand{\nn}{\nonumber}

\newcommand{\be}{\begin{equation}} 
\newcommand{\ee}{\end{equation}} 
\newcommand{\ba}{\begin{eqnarray}} 
\newcommand{\ea}{\end{eqnarray}}

\newcommand{\pa}{\partial}

\usepackage{dcolumn}

\begin{document}
                                                                                
\title{Open bottom states and the $\bar B$-meson propagation in hadronic matter}
\author{Juan M. Torres-Rincon}
\affiliation{Institut de Ci\`encies de l'Espai (IEEC/CSIC), Campus Universitat Aut\`onoma
de Barcelona, Facultat de Ci\`encies, Torre C5, E-08193 Bellaterra, Spain}
\affiliation{Subatech, UMR 6457, IN2P3/CNRS, Universit\'e de Nantes, \'Ecole de Mines de Nantes, 4 rue Alfred Kastler 44307,
Nantes, France}
\author{Laura Tolos}
\affiliation{Institut de Ci\`encies de l'Espai (IEEC/CSIC), Campus Universitat Aut\`onoma
de Barcelona, Facultat de Ci\`encies, Torre C5, E-08193 Bellaterra, Spain}
\affiliation{Frankfurt Institute for Advanced Studies. Johann Wolfgang Goethe University, Ruth-Moufang-Str. 1,
60438 Frankfurt am Main, Germany }
\author{Olena Romanets}
\affiliation{KVI, University of Groningen, Zernikelaan 25, 9747AA Groningen, The Netherlands}
\date{\today}


\begin{abstract}
The interaction and propagation of $\bar B$ mesons with light mesons, $N$ and $\Delta$ is studied within a unitarized approach
based on effective models that are compatible with chiral and heavy-quark symmetries. We find several heavy-quark spin doublets
in the open-bottom sectors, where $\bar B$ and $\bar B^*$ mesons are present. 
In the meson sector we find several resonant states, among them, a $B_0$ and a $B_1$ with masses 5530 MeV and 5579 MeV as well as $B_{s0}^*$ and $B_{s1}^*$ narrow states at 5748 MeV and 5799 MeV, respectively.
They form two doublets with no experimental identification yet, the first one being the bottom counterpart of the $D_0(2400)$ and $D_1(2430)$ states, and the second bottom doublet associated to
the ubiquitous $D_{s0}^* (2317)$ and the $D_{s1} (2460)$.
In the baryon sector, several $\Lambda_b$ and $\Sigma_b$ doublets are identified, among them the one given by the experimental $\Lambda_b(5910)$ and
$\Lambda^*_b(5921)$. Moreover, one of our states, the $\Sigma_b^*(5904)$, turns out to be the  bottom counterpart of the $\Sigma^*(1670)$ and
$\Sigma_c^*(2549)$, which is a case for discovery. We finally analyze different transport coefficients for the $\bar B$ meson in hot matter,
such as formed in heavy-ion collisions at RHIC and LHC. For RHIC/LHC energies, the main contribution to the coefficients comes
from the interaction of $\bar B$ mesons with pions. However, we also include the effects of baryonic density which might be sizable at temperatures $T \lesssim$ 100 MeV, as the chemical potential
is expected to increase in the last stages of the expansion. We conclude that although the relaxation time decreases with larger
baryonic densities, the $\bar B$ meson does not thermalize at RHIC/LHC energies, representing an ideal probe for the initial bottom distribution.
\end{abstract}

\maketitle

\section{Introduction}

The properties of matter created in heavy-ion collisions (HICs) have been a subject of interest over the past decades. Most of the studies have been focused in the
potential signatures and features of the deconfined phase, the quark-gluon plasma (QGP).  For the characterization of this phase, hadrons
with heavy flavor (charm or bottom) play a fundamental role as heavy quarks, produced in the early stage of the collision, can probe the formed medium
during its entire evolution. When the medium cools down, the hadronization takes place and, after freeze-out, heavy-flavored hadrons are finally detected~\cite{Abelev:2012qh}. 
Therefore, heavy hadrons --such as $D$ and $B$ mesons-- are considered to be an efficient and unique probe for testing the hot and dense medium created in HICs, in both QGP and hadronic phases. 

Focusing on the latter, the diffusion of $D$ mesons in hadronic matter was initially obtained within an effective theory that incorporates both chiral and heavy-quark
symmetries~\cite{Laine:2011is} and also using parametrized interactions with light mesons and baryons~\cite{He:2011yi}. Moreover, effective Lagrangians at leading order were used to obtain the
scattering amplitudes of $D$ mesons with light mesons and baryons~\cite{Ghosh:2011bw}. However, the need of unitarization was later pointed out in order to avoid unphysical large transport
coefficients~\cite{Abreu:2011ic} and also next-to-leading order contributions were considered~\cite{Abreu:2011ic,Tolos:2013kva,Torres-Rincon:2013nfa}.

More recently, the propagation of bottom mesons in matter, such as $B$ mesons has been analyzed. The drag and diffusion coefficients of open bottom mesons in a hadronic medium of pions, kaons and
etas was evaluated with the use of scattering lengths as dynamical input~\cite{Das:2011vba}. In that work, inelastic channels and non-perturbative processes were not taken into account. The non-perturbative
character of the $B$ meson interaction in a mesonic environment was addressed in Ref.~\cite{Abreu:2012et}, and it was found to be relevant for the determination of the transport coefficients, as in the case of $D$ mesons.

In this paper we aim at, first, analyzing the scattering of $\bar B$ mesons with light mesons and baryons, such as $N$ and $\Delta$, within a unitarized approach in coupled channels taking, as bare interaction, effective
models that are compatible with chiral and heavy-quark symmetries. Note that in this paper we study the $\bar B$ meson, the counterpart of the $D$ meson in the bottom sector. In the meson sector, we extend
the results of Ref.~\cite{Abreu:2012et} including the coupled-channel structure of the interaction of $\bar B$ with pions, kaons, anti-kaons and etas (by also incorporating the interaction of $B_s$ with light mesons in the coupled channel structure).  In the
baryonic sector, we continue the study initiated in Ref.~\cite{GarciaRecio:2012db} with regard to the open-bottom baryon states in order to determine the scattering of $\bar B$ mesons with $N$ and $\Delta$. We then obtain
the transport coefficients for a $\bar B$ meson in this hadronic environment by making use of our previous knowledge of the scattering of $\bar B$ mesons in matter. We present results for the diffusion and drag 
coefficients at zero baryochemical potential which can be used in transport calculations for high-energy collisions, such as those at the Relativistic Heavy Ion Coolider (RHIC)~\cite{rhic} or the Large Hadron Collider (LHC)~\cite{lhc}.  Afterwards, we discuss the behavior of
the relaxation time and the spatial diffusion coefficient for isentropic trajectories ranging from RHIC/LHC conditions to FAIR~\cite{fair} at its top energy.

The paper is organized as follows. In Sec.~\ref{sec:interaction} we study the interaction of $\bar B$ mesons with light mesons, nucleons and $\Delta$ within unitarized effective theories and present the dynamically-generated open bottom states. In
Sec.~\ref{sec:transport} we introduce the relevant transport coefficients for heavy mesons and present our results as a function of temperature, baryochemical potential (or entropy per baryon) and the momentum of the heavy meson. Our conclusions are given in Sec.~\ref{sec:conclusions}.

\section{Open bottom states \label{sec:interaction}}

In this section we study the interaction of $\bar B$ mesons with hadrons within a unitarized approach in coupled-channels based on effective models that are compatible with chiral and heavy quark symmetries, in
particular heavy-quark spin symmetry (HQSS). The unitarization in coupled channels has proven to be very successful in describing some of the existing experimental data on baryon and meson states as dynamically
generated states. These are obtained as poles of the scattering amplitudes in coupled-channel basis, that is  usually characterized by different quantum numbers, such as bottom ($B$), charm ($C$), strange ($S$),
isospin ($I$) and spin ($J$). We concentrate on the sectors with $B=-1$ and $C=0$, where $\bar B$ (and also $\bar B^*$) mesons interact with light mesons as well as  $N$ and $\Delta$ baryons.

The scattering amplitudes $T$ for the interaction of $\bar B$ mesons with light mesons and baryons follows the standard multichannel scattering (integral) Bethe-Salpeter (BS) equation,
\be
T= V + V  G T \ ,
\label{bs}
\ee
where $V$ is the potential resulting from the meson-meson (baryon-meson) effective Lagrangian and $G$ is the two-particle meson-meson (baryon-meson) propagator. 

The kernel $V$ is a matrix that consists of all possible meson-meson (baryon-meson) transitions. We focus on the interaction of $\bar B$ mesons with the
pseudo-Goldstone bosons ($\pi$, $K$, $\bar K$ and $\eta$) as well as with the lightest baryons ($N$ and $\Delta$). We make use of the effective
model of Ref.~\cite{Abreu:2011ic,Abreu:2012et} for the interaction of $\bar B$ mesons with light mesons, which is consistent with chiral and heavy-quark symmetries. For
the scattering of $\bar B$ mesons with  baryons, we take into account the SU(6)$\times$HQSS WT scheme of Refs.~\cite{GarciaRecio:2008dp,Gamermann:2010zz,Romanets:2012hm,GarciaRecio:2012db,Garcia-Recio:2013gaa}. 
Similarly to the meson-meson sector, the baryon-meson model fulfills chiral symmetry in the light-quark sector while heavy-quark symmetry constraints are respected in the heavy-quark sector. The details of these
effective models will be given in the Secs.~\ref{sec:mes} and \ref{sec:bar}.

The $V$ kernel can be factorized in the on-mass shell~\cite{Oller:1997ti}, so the scattering amplitudes $T$ of Eq.~(\ref{bs}) are the solutions of a set of linear algebraic coupled equations 
\be
T_{ij}=[1 -  V G ]_{ik}^{-1} \ V_{kj} \ ,
\ee
where $i$ and $j$ indicate the initial meson-meson (baryon-meson) and final meson-meson (baryon-meson) systems, respectively. This approach is practically equivalent to
the so-called $N/D$ method \cite{Oller:2000fj}. In the on-shell {\it ansatz}, the two-particle propagators ---often called loop functions--- form a diagonal matrix $G$. The loop
function reads
\be \label{eq:loop} G_r (\sqrt{s})= i \gamma_r \int \frac{d^4q}{(2\pi)^4} \frac{1}{(P-q)^2-M_r^2+i\epsilon} \frac{1}{q^2-m_r^2+i\epsilon} \ , \ee
with the total four-momentum $P$ related to the center-of-mass  (C.M.) squared energy $s$ by $s=P^2$,  and $q$ being the relative four-momentum in the center-of-mass frame. 
The quantities $m_r$ and $M_r$ stand for the masses of the two particles propagating in the intermediate channel $r$, i.e,  two mesons, or a meson and a baryon. The
factor $\gamma_r$ has been introduced to account for the different normalization of the meson-meson and baryon-meson interactions. In fact, as we will see
in the following subsections, $\gamma_r=1$ for the adimensional meson-meson $V$ kernel while for the baryon-meson sector $\gamma_r = 2 M_r$, with $M_r$ being the mass
of the baryon. The meson-meson (baryon-meson) loop functions are divergent and are regularized by means of dimensional regularization.

In order to study the dynamically-generated resonances, we study both
the first and second Riemann sheets of the C.M. energy $\sqrt{s}$. The poles of the
scattering amplitude on the first Riemann sheet that appear on the real
axis below threshold are interpreted as bound states. The poles that are found
on the second Riemann sheet below the real axis and above threshold are
identified with resonances. Note that we often refer to all poles generically as
resonances, regardless of their concrete nature, since usually they can decay
through other channels not included in the model space. The mass and the
width of the bound state/resonance can be found from the position of the pole
on the complex energy plane. Close to the pole, the $T$-matrix behaves
as 
\begin{equation} \label{Tfit} T_{ij}  (s) \approx \frac{g_i
e^{\rm{i}\phi_i}\,g_je^{\rm{i}\phi_j}}{z(s)-z_R} \,.  \end{equation} %

where, in the baryon-meson sector,  $z(s)=\sqrt{s}$  and $z_R=M_R - \rm{i}\, \Gamma_R/2$ provides the mass ($M_R$) and the width
($\Gamma_R$) of the resonance, while $g_j e^{{\rm i}\phi_j}$ (modulus and phase) is
the (adimensional) coupling of the resonance to the channel $j$. In the usual parametrization for the meson-meson scattering, $z(s)=s$ and
$z_R$ is the pole position in the $s$ plane with a coupling with dimensions of energy.

\begin{table*}
\begin{ruledtabular}
\begin{tabular}{cccccc}
$(S,I)$ & Channel & $C_0 $ & $C_1$ & $C_2$ & $C_3$ \\
\hline
$(0, \frac{1}{2})$ & $\bar B \pi \rightarrow \bar B \pi$ & $-2$ & $-3 m_\pi^2$ & $1$ & $1$ \\
 &  $\bar B \pi \rightarrow \bar B \eta$ & 0 & $-3 m_\pi^2$ & $1$ & $1$ \\
 &  $\bar B \eta \rightarrow \bar B \eta$ & 0 & $-m_\pi^2$ & $1/3$ & $1/3$ \\
 &  $\bar B_s \bar K \rightarrow \bar B_s \bar K$ & $-1$ & $-3m_K^2$ & $1$ & $1$ \\
 &  $\bar B \pi \rightarrow \bar B_s \bar K$ & $-\sqrt{6}/2$ & $-3\sqrt{6} (m_K^2+m^2_\pi) /4$ & $\sqrt{6}/2$ & $\sqrt{6}/2$ \\
 &  $\bar B \eta \rightarrow \bar B_s \bar K$ & $-\sqrt{6}/2$ & $\sqrt{6} (5m_K^2-3m^2_\pi) /4$ & $-\sqrt{6}/6$ & $-\sqrt{6}/6$ \\
 $(0, \frac{3}{2})$ & $\bar B \pi \rightarrow \bar B \pi$ & $1$ & $-3 m_\pi^2$ & $1$ & $1$ \\
 $(1, 0)$ & $\bar B K \rightarrow \bar B K$ & $-2$ & $-6 m_K^2$ & $2$ & $2$ \\
  & $\bar B_s \eta \rightarrow \bar B_s \eta$ & $0$ & $-2 (3m^2_\eta-m^2_\pi)$ & $4/3$ & $4/3$ \\
  & $\bar B K \rightarrow \bar B_s \eta$ & $-\sqrt{3}$ & $-\sqrt{3} (5m^2_K-3m^2_\pi)/2$ & $\sqrt{3}/3$ & $\sqrt{3}/3$ \\
    $(1, 1)$ & $\bar B K \rightarrow \bar B K$ & $0$ & $0$ & $0$ & $0$ \\
  & $\bar B_s \pi \rightarrow \bar B_s \pi$ & $0$ & $0$ & $0$ & $0$ \\
  & $\bar B K \rightarrow \bar B_s \pi$ & $1$ & $-3 (m^2_K+m^2_\pi)/2$ & $1$ & $1$ \\
    $(-1, 0)$ & $\bar B \bar K \rightarrow \bar B \bar K$ & $-1$ & $3m_K^2$ & $-1$ & $-1$ \\
   $(-1, 1)$ & $\bar B \bar K \rightarrow \bar B \bar K$ & $1$ & $3m_K^2$ & $1$ & $1$  \\
   $(2, \frac{1}{2})$ & $\bar B_s K \rightarrow \bar B_s K$ & $1$ & $-3m_K^2$ & $1$ & $1$  \\
\end{tabular}
\end{ruledtabular}
\caption{ \label{tab:isoscoeff} Isospin coefficients of the scattering amplitudes for the $\bar B$ meson--light meson channels with total strangeness $S$ and isospin $I$. }
\end{table*}

\subsection{Bottom meson resonances}
\label{sec:mes}

The interaction between the $\bar B$ mesons and the pseudoscalar Goldstone bosons is given by the effective Lagrangian in Refs~\cite{Lutz:2007sk,Guo:2008gp,Guo:2009ct,
Geng:2010vw,Abreu:2011ic}. In particular we adapt the $B$-meson interaction from our past work~\cite{Abreu:2012et} to the present case,
where the $\bar B$ field is given by $\bar B=(B^-,\bar B^0,\bar B_s^0)$.  

At leading-order (LO) in heavy-quark mass expansion and next-to-leading order (NLO) in the chiral expansion the tree-level scattering amplitude of a $\bar B$ meson interacting with light mesons reads
\begin{eqnarray} \label{eq:potmeson} V^{IJSB} &=& \frac{C_0}{4 f_\pi^2} (s-u) + \frac{2C_1 h_1}{3 f_\pi^2}+ \frac{2C_2}{f_\pi^2} h_3 (p_2 \cdot p_4) 
 \\ & + & \frac{2C_3}{f_\pi^2} h_5 [(p_1 \cdot p_2) (p_3 \cdot p_4)
+ (p_1 \cdot p_4) (p_2 \cdot p_3)] \nn , \end{eqnarray}
where $p_1$ and $p_2$ are the four-momenta of the incoming hadrons, $p_3$ and $p_4$ the outgoing momenta, and $s=(p_1 + p_2)^2$ and $u=(p_1 - p_4)^2$.  At LO in the heavy-quark expansion, the scattering amplitude for $\bar B^*$ meson with light mesons coincides (modulus the
polarization vectors) with the amplitude of Eq.~(\ref{eq:potmeson}). For completeness, we will thus also show the results in the $J=1$ channel, with the only heavy-quark breaking effect being the physical masses of the bottom mesons.

The quantities $C_i$ are the isospin coefficients of the different scattering amplitudes of $\bar B$ mesons with $\pi$, $K$, $\bar K$ and $\eta$ mesons, which are shown
in Table~\ref{tab:isoscoeff}.  The $h_i$ coefficients are the low-energy constants (LECs). We fix $h_1=-1.042$ using the mass difference between the $B$ and $B_s$ mesons~\cite{Geng:2010vw}, 
whereas $h_3$ and $h_5$ are free.  With the inclusion of all coupled channels and the analysis of the scattering amplitudes in the whole complex plane we have found that the previously used values of $h_3$ and $h_5$ provided a too large NLO contribution with respect to LO.  We recalibrate $h_3$ and $h_5$ keeping a more conservative
(smaller) values. The numbers we use are $h_3=0.25$ and $h_5=-0.015$ GeV$^{-2}$. 
In order to solve the BS equation of Eq.~(\ref{bs}), the loop function needs to be renormalized. We keep the prescription of Ref.~\cite{Abreu:2012et}, which consists
on fixing the value of the loop function in dimensional regularization at $\mu=1 \ {\rm GeV}$ to the one coming from cutoff regularization for $\Lambda=770$ MeV at the energy threshold 
of the lightest channel, $m_B+m_\pi$. In this case, the subtraction constant is set to $a(\mu)=-3.38$. 
In fact, as we shall see, the combination of the free LECs and the subtraction constant are determined to reproduce a state, $B_0$,  with a similar mass of that  found in Ref.~\cite{Abreu:2012et}.

\subsubsection{$B$ states ($J=0$)}

\begin{table*}
\begin{ruledtabular}
\begin{center}
\begin{tabular}{  c  c  c  c  c  }
 $M_{R}$ & $\Gamma_{R}$   & Couplings  & $(S,I)$	& Open \\
 (MeV) & (MeV)  &   to main channels (MeV$^{1/2}$) & & channels \\
 \hline
5530.3  &  238.5  &   $g_{\bar B \pi}=25.3$,  $g_{\bar B \eta}=2.5$,  $g_{\bar B_s \bar K}=11.5$  &  {(0,1/2)} & $\bar B \pi$ \\
5827.0  &  48.1   &   $g_{\bar B \pi}=6.7$,  $g_{\bar B \eta}=16.1$, $g_{\bar B_s \bar K}=26.2$            &  {(0,1/2)} & $\bar B \pi, \bar B \eta$ \\
5747.6  &  0.0    &   $g_{\bar B K}=19.9$,  $g_{\bar B_s \eta}=13.9$                                 &  {(1,0)}   &  \\
5774.0  &  0.2    &   $g_{\bar B \bar K}=7.1$                                                              &  {(-1,0)}  & $\bar B \bar K$ \\
\end{tabular}
\end{center}
\end{ruledtabular}
\caption{Masses, widths and couplings to meson-meson channels of the 
$\bar B$ resonances $(J=0)$. In the first and second column we present the mass and width of 
these states, respectively. The next column displays the (modulus of the) couplings to
the different meson-meson channels, ordered by threshold energies. The fourth column
indicates the strangeness ($S$) and isospin ($I$) of the resonance while in the last column we show the meson-meson channels that are allowed for decay.
}
\label{tab:mesonJ0}
\end{table*}

\begin{table*}
\begin{ruledtabular}
\begin{center}
\begin{tabular}{ c  c  c  c  c }
 $M_{R}$ & $\Gamma_{R}$   & Couplings  & $(S,I)$	& Open \\
 (MeV) & (MeV)  &   to main channels (MeV$^{1/2}$) & & channels \\
 \hline
5579.2  &  251.9  &   $g_{\bar B^* \pi}=25.9$,  $g_{\bar B^* \eta}=2.9$, $g_{\bar B^*_s \bar K}=12.0$    &  {(0,1/2)} & $\bar B^* \pi$ \\
5880.4  &  53.0   &   $g_{\bar B^* \pi}=6.5$,  $g_{\bar B^* \eta}=15.3$, $g_{\bar B^*_s \bar K}=25.1$  &  {(0,1/2)} & $\bar B^* \pi, \bar B^* \eta$ \\
5798.7  &  0.0    &   $g_{\bar B^* K}=19.3$,  $g_{\bar B^*_s \eta}=13.9$                                    &  {(1,0)}   & \\
5820.0  &  0.7  &   $g_{\bar B^* \bar K}=9.8$                                                                   &  {(-1,0)}  & $\bar B^* \bar K$ \\
\end{tabular}
\end{center}
\end{ruledtabular}
\caption{As in Table~\ref{tab:mesonJ0}, but for $B^*$ meson resonances ($J=1$).
}
\label{tab:mesonJ1}
\end{table*}

In Table~\ref{tab:mesonJ0} we show the mass and width of the different $J=0$ states in the $B=-1$ sector together with their couplings to the different meson-meson channels and 
the meson-meson channels that are allowed for decay.  The resonance in the $(S,I)$=(0,1/2) sector at 5530 MeV is assigned to a wide $B_0$ resonance (not yet experimentally seen) in analogy
to the experimental $D_0(2400)$ in the charm sector \cite{pdg}.\footnote{The amplitudes
 of $B$ and $\bar B$ mesons interacting with light mesons at LO in
the heavy-quark mass expansion and NLO in the chiral expansion are related by charge conjugation.}.

In Ref.~\cite{Guo:2006fu} this resonant state is seen at $5536$ MeV with a width of $234$ MeV using a similar method at LO in the chiral expansion. Within the non-linear chiral SU(3) model of~\cite{Kolomeitsev:2003ac}, this state
is located at 5526 MeV but no width is provided. We also observe a second narrower resonance in $(S,I)=(0, 1/2)$ at 5827 MeV, which was overlooked in~\cite{Abreu:2012et}. This state is identified 
in \cite{Guo:2006fu} with $M_R=5842$ MeV and $\Gamma=35$ MeV, and with  $M_R=5760$ MeV and a width of approximately 30 MeV in \cite{Kolomeitsev:2003ac}. Moreover, we find two narrow states. The first one at
5748 MeV is seen in the $(1,0)$ channel, that mainly couples to $\bar B K$ channel. 
In Ref.~\cite{Guo:2006fu} it is located at 5729 MeV while in Ref.~\cite{Kolomeitsev:2003ac}  it is found at 5643 MeV. The state in the $(S,I)=(-1,0)$ channel lies at 5774 MeV, close to the
bound state found in~\cite{Kolomeitsev:2003ac}. No state is seen in the $(S,I)=(1,1)$ channel, in contrast to the findings of~\cite{Kolomeitsev:2003ac}.

\subsubsection{$B^*$ states ($J=1$)}

We show in Table~\ref{tab:mesonJ1} the $J=1$  states. Two wide resonances are found with masses 5579 MeV and 5880 MeV that couple strongly to $\bar B^* \pi$ and $\bar B_s^* \bar K$, respectively. The
first one is the charm counterpart of the $D_1(2430)$ state. Furthermore, two narrow states at 5799 MeV and 5820 MeV are seen, with a strong coupling to $\bar B^* \bar K$.

In Ref.~\cite{Kolomeitsev:2003ac} is also found that the $1^+$ spectrum resembles the $0^+$ sector, predicting states at 5590 MeV and 5810 MeV $(S,I)=(0,1/2)$, 5690 MeV for $(S,I)=(1,0)$, 5807 MeV in $(S,I)=(-1,0)$, and 5790 MeV in $(S,I)=(1,1)$.
We generate similar states to those reported in Ref.~\cite{Kolomeitsev:2003ac},  with the exception of the resonance in the $(S,I)=(1,1)$ sector. In Ref.~\cite{Guo:2006rp}, a bound state with mass of 5778 MeV was
obtained in the $(S,I)=(1,0)$ sector, similar to our bound state at 5799 MeV. In the $(S,I)=(1/2,0)$ channel, two  states were found in~\cite{Guo:2006rp} with masses similar to ours. The $(S,I)=(-1,0)$ sector was not explored in Ref.~\cite{Guo:2006rp}.

Note that at LO in heavy-quark expansion, the $J=0$ and $J=1$ sectors are decoupled \cite{Abreu:2012et} and that an analogous set of states to the $J=0$  sector is obtained
due to HQSS.  In fact, the $J=0$ and $J=1$ states form HQSS doublets.  We define a HQSS doublet as a pair of $J=0$ and $J=1$ states that are degenerate when HQSS is restored. Such states have similar masses, with the $J=0$ state
coupling strongly to a two-particle channel with one of the intervening particles being the HQSS partner of one of the particles in the dominant two-particle channel for the generation
of the $J=1$ state. This is the case, for example, of the $B_0(5530)$ and $B_1(5579)$, which turn out to be the bottom counterparts of the experimental $D_0(2400)$ and $D_1(2430)$, as well
as the $B_{s0}^*(5748)$ and $B_{s1}^*(5799)$, these last two possibly being  the bottom homologues of the $D^*_{s0}(2317)$ and the $D_{s1} (2460)$ states, respectively.

\subsection{Bottom baryon resonances}
\label{sec:bar}

We  follow here the approach applied in  Refs.~\cite{GarciaRecio:2008dp,Gamermann:2010zz,Romanets:2012hm,Garcia-Recio:2013gaa} for charm quarks and recently used in the bottom sector \cite{GarciaRecio:2012db}. The model obeys SU(6) spin-flavor symmetry and also HQSS  \cite{Garcia-Recio:2013gaa}. This is a model extension of the WT SU(3) chiral Lagrangian \cite{GarciaRecio:2008dp,Romanets:2012hm}. The extended SU(6)$\times$HQSS WT baryon-meson interaction is given by

\begin{eqnarray}
V_{ij}^{IJSB}(s) & = & \frac{D_{ij}^{IJSB}}{4\,f_i f_j}
(2\sqrt{s}-M_i-M_j) \nonumber \\ & \times &\sqrt{\frac{M_i+E_i}{2M_i}}
\sqrt{\frac{M_j+E_j}{2M_j}} 
\,.
\label{eq:vsu8break}
\end{eqnarray}

The $i$ ($j$) are the outgoing (incoming) baryon-meson channels while $M_i$, $E_i$ and $f_i$ stand for 
the baryon mass and energy, in the C.M. frame, and the meson decay constant in the $i$ channel, respectively. The
masses of baryons with bottom content used in this work are compiled in Tables I of Ref.~\cite{GarciaRecio:2012db}, while those of the bottom mesons and their decay
constants are given in Table II of Ref.~\cite{GarciaRecio:2012db}. The rest of hadron masses
and meson decay constants  have been taken from
Ref.~\cite{Romanets:2012hm}. The $D_{ij}^{IJSB}$ elements  are the coefficients coming
 from the underlying SU(8) group structure in the Appendix B of
Ref.~\cite{Romanets:2012hm}, where one can identify the charm $C=1$
sector couplings given there with those needed here that correspond to
the $B=-1$ sector.

In order to solve the BS equation of Eq.~(\ref{bs}), the loop function is renormalized by a subtraction constant such that %
\begin{equation}
G_{r}^{IJS}=0 \quad\text{at~~} \sqrt{s}=\mu^{IS}. 
\label{eq:musi}
\end{equation}
To fix the subtraction point $\mu^{IS}$ we apply the following prescription:
 $\mu^{IS}$ is independent of $J$ and is taken as
$\sqrt{m_{\rm{th}}^2+M_{\rm{th}}^2}$, where $m_{\rm{th}}$ and $M_{\rm{th}}$,
are respectively, the masses of the meson and baryon producing the lowest
threshold (minimal value of $m_{\rm{th}}+M_{\rm{th}}$).

We concentrate on all $B=-1$ sectors where $\bar B$ mesons interact with $N$ and $\Delta$ since we are interested in studying the propagation of $\bar B$ in a hadronic
environment. These are the $\Lambda_b$ and $\Lambda_b^*$ ($I=0; J=1/2,3/2$), the $\Sigma_b$ and $\Sigma_b^*$ ($I=1; J=1/2,3/2$) and the ($I=2; J=3/2$) sectors. Note that the
vacuum $\Delta$-decay width has to be considered for the determination of the dynamically-generated resonances. This effect is introduced in the unitarization procedure through
a convolution of the $\bar B$ $\Delta$ propagator with the corresponding spectral function of the $\Delta$ baryon, as done in Ref.~\cite{Gamermann:2010zz}. Only the resonances that lie
close to the $\bar B$ $\Delta$ channel, as compared to the $\Delta$ width, and that couple strongly to this system will be affected. 

\subsubsection{$\Lambda_b$ and $\Lambda_b^*$ states \rm{($I=0; J=1/2,3/2$)}}

\begin{table*}
\begin{ruledtabular}
\begin{center}
\begin{tabular}{  c  c  c c  c  }
 $M_{R}$ & $\Gamma_{R}$   & Couplings  & $J$	& Open \\
 (MeV) & (MeV)  &   to main
 channels &  &channels \\ 

\hline
 5797.6 & 0.0 &   ${g_{N \bar B}=4.9}$,   $g_{N \bar B^*}=8.3$,  
       $g_{\Lambda \bar B_s^{0}}=2.1$,    $g_{\Lambda \bar B_s^{*}}=3.6$
   &       $1/2$  & \\

 {5910.1}  & 0.0 &  $g_{\Sigma_b \pi}=1.8$,   ${ g_{N \bar B}}=4.6$,
    $g_{N \bar B^*}=3.0$,    $g_{\Lambda_b \omega}=1.4$       & {{1/2}}  & \\

  {{5921.5}} &  0.0  &  $g_{\Sigma_b^* \pi}=1.8$,   $g_{N \bar B^*}=5.7$, 
    $g_{\Lambda_b \omega}=1.5$                      & {{3/2}}   & \\

   6009.3  &  0.0 &        $g_{\Lambda_b \eta}=2.0$,   
    $g_{N \bar B^*}=1.7$, $g_{\Lambda \bar B_s^{0}}=3.9$,  $g_{\Lambda \bar B_s^{*}}=6.0$   & 1/2  & $\Sigma_b \pi$ \\  

{{6034.0}}  &  4.7  &   ${g_{N \bar B}=3.2}$,  $g_{N \bar B^*}=2.2$,
  $g_{\Sigma_b \rho}=2.2$,  $g_{\Sigma_b^* \rho}=1.4$   &  {{1/2}} & $\Sigma_b \pi$ \\

 {{6044.8}} &  4.0  &   $g_{N \bar B^*}=4.$,    $g_{\Lambda \bar B_s^*}=1.3$,  $g_{\Sigma_b \rho}=1.1$,
 $g_{\Sigma_b^* \rho}=2.4$ &  {{3/2}} & $\Sigma_b^* \pi$\\

6090.8  & 0.0 &  $g_{N \bar B^*}=1.$,  $g_{\Xi_b K}=2.$,  $g_{\Lambda \bar B_s^{*}}=1.2$,
$g_{\Sigma_b^* \rho}=1.2$ & 1/2 & $\Sigma_b \pi$ \\

{{6094.1}}  &  2.6  &  $g_{\Xi'_b K}=1.7$,  $g_{\Lambda \bar B_s^{0}}=5.7$,  $g_{\Lambda \bar B_s^{*}}= 3.8$, 
 $g_{\Lambda_b \phi}=1.4$ & {{1/2}} & $\Sigma_b \pi$ \\

 {{6105.4}} & 2.5 &  $g_{\Xi_b^* K}=1.7$,  
 $g_{\Lambda \bar B_s^*}=7.1$,  $g_{\Lambda_b \phi}=1.4$, 
 $g_{\Xi_b^* K^*}=1.6$  & {{3/2}} & $\Sigma_b^* \pi$\\

{{6201.9}}  &  54.3  &   $g_{\Lambda_b \omega}=2.2$,  
  $g_{\Lambda \bar B_s^{0}}=0.7$,    $g_{\Sigma_b \rho}=1.1$,  $g_{\Xi'_b K^*}=0.7$   &  {{1/2}} & $\Sigma_b \pi$, $\Lambda_b \eta$ \\

{{6207.5}}  &  54.2  &   $g_{\Lambda_b \omega}=2.2$,  $g_{\Lambda \bar B_s^*}=0.9$,  
 $g_{\Sigma_b^* \rho}=1.1$,  $g_{\Xi_b^* K^*}=0.8$  &  {{3/2}} & $\Sigma_b^* \pi$  \\

6243.4  &  19.5  &      
$g_{\Xi_b K}=1.0$,   $g_{\Sigma_b \rho}=1.4$,  $g_{\Sigma_b^* \rho}=2.1$  
&   $1/2$ & $\Sigma_b \pi$, $\Lambda_b \eta$, $N \bar B$ \\

{{6361.9}}  & 0.1 &   $g_{\Xi'_b K}=1.6$, $g_{\Lambda_b \phi}=1.5$
 &       ${{1/2}}$ & $\Sigma_b \pi$, $\Lambda_b \eta$ \\
 & & & & $N \bar B$, $N \bar B^*$, $\Xi_b K$\\

{{6373.3}} &  0.1    &   $g_{\Xi_b^* K}=1.6$,
$g_{\Lambda \bar B_s^*}=1.0$,  $g_{\Lambda_b \phi}=1.5$
& {{3/2}} & $\Sigma_b^* \pi$, $N \bar B^*$ \\

6403.9  &  45.9  &  $g_{\Xi_b K}=0.7$,   
$g_{\Xi'_b K^*}=1.8$, $g_{\Xi_b^* K^*}=2.4$
  &  1/2  & $\Sigma_b \pi$, $\Lambda_b \eta$, $N \bar B$ \\
 &  &     & & $N \bar B^*$, $\Xi_b K$, $\Lambda_b \omega$ \\

6459.0  &  0.06  &   $g_{\Sigma_b \rho}=2.3$,  $g_{\Sigma_b^* \rho}=1.0$
& 3/2  & $\Sigma_b^* \pi$, $N \bar B^*$ \\
& & & &$\Lambda_b \omega$, $\Xi_b^* K$ \\

{{6463.8}} & 1.6 &    $g_{\Lambda_b \phi}=1.6$,   $g_{\Xi_b K^*}=2.3$
       & {{1/2}} & $\Sigma_b \pi$, $\Lambda_b \eta$,   $N \bar B$ \\
 & &     &   & $N \bar B^*$, $\Xi_b K$, $\Lambda_b \omega$, $\Xi'_b K$\\

{{6464.4}}  &  1.4  &  $g_{\Lambda_b \phi}=1.5$,   $g_{\Xi_b K^*}=2.3$
 &  {{3/2}}  & $\Sigma_b^* \pi$, $N \bar B^*$\\
 &&&& $\Lambda_b \omega$, $\Xi_b^* K$ \\

{{6515.6}}  &  6.1   &    $g_{\Lambda_b \phi}=1.1$,  $g_{\Xi'_b K^*}=1.8$, 
$g_{\Xi_b^* K^*}=1.4$   
     &  {{1/2}}  & $\Sigma_b \pi$, $\Lambda_b \eta$, $N \bar B$,
$N \bar B^*$ \\
  & &    & & $\Xi_b K$,  $\Lambda_b \omega$,
$\Xi'_b K$, $\Lambda \bar B_s^{0}$\\

{{6520.2}}  &  6.2   &    $g_{\Lambda_b \phi}=1.0$,    $g_{\Xi'_b K^*}=1.1$,
$g_{\Xi_b^* K^*}=2.1$
 &  {{3/2}}  & $\Sigma_b^* \pi$, $N \bar B^*$\\
 &&&&  $\Lambda_b \omega$, $\Xi_b^* K$\\

6590.7  &  0.02  &   
$g_{\Xi'_b K^*}=2.5$, $g_{\Xi_b^* K^*}=1.3$
& 3/2  & $\Sigma_b^* \pi$, $N \bar B^*$, $\Lambda_b \omega$\\
 &  &      & & $\Xi_b^* K$, $\Lambda \bar B_s^*$, $\Sigma_b \rho$
\\

\end{tabular}
\end{center}
\end{ruledtabular}
\caption{Masses, widths and the most important couplings of the 
$\Lambda_b$ and $\Lambda_b^*$ baryon resonances $(I=0; J=1/2, J=3/2)$. In the first and second column we present the mass and width of these states. The next column displays the (modulus of the) dominant couplings to the different baryon-meson channels, ordered by the threshold energies. The fourth column indicates the spin of the resonance whereas in the last column we show the baryon-meson channels that are allowed for decay.}
\label{TabPoles0}
\end{table*}

In the $\Lambda_b$ sector,
 the following sixteen channels are
involved:
\noindent
\begin{center}
\begin{tabular}{llllllll}
$\Sigma_b \pi $   & $\Lambda_b \eta$   & $N \bar B$  & $N \bar B^*$    
& $\Xi_b K$    &  $\Lambda_b \omega$     &  $ \Xi'_b K$  & $ \Lambda \bar B_s^{0} $   \\
$ \Lambda \bar B_s^{*}$  &   $\Lambda_b \eta'$    &  $\Sigma_b \rho$    &  $\Sigma_b^* \rho$   & $\Lambda_b \phi$    
& $\Xi_b K^*$    & $\Xi'_b K^*$     & $\Xi_b^* K^*$  \\
\end{tabular}
\end{center}
%
Likewise for the $\Lambda^*_b$ sector,
there are eleven channels:
\noindent
\begin{center}
\begin{tabular}{llllll}
$ \Sigma_b^* \pi $   & $ N \bar B^*  $  & $ \Lambda_b \omega  $   & $ \Xi_b^* K  $    
& $ \Lambda \bar B_s^*  $    & $ \Sigma_b \rho  $  \\
 $ \Sigma_b^* \rho  $  & $ \Lambda_b \phi  $   & $ \Xi_b K^* $ 
& $ \Xi'_b K^*  $   & $  \Xi_b^* K^*  $ &  \\
\end{tabular}
\end{center}
In both cases the channels are ordered by increasing mass thresholds.

In Table \ref{TabPoles0} we show the $J=1/2$ and $J=3/2$ dynamically-generated states ordered by increasing mass. In the first and second columns we present the masses and widths
of these states. The next column displays the (modulus of the) couplings to the different dominant baryon-meson channels, ordered by the threshold energies. The fourth column indicates the spin of the resonance while in the last column we show the baryon-meson channels that are allowed for decay. 

Results on the $\Lambda_b$ and $\Lambda_b^*$ sectors have been previously discussed in Ref.~\cite{GarciaRecio:2012db}. However, in this latter work only the states coming from the most attractive SU(8) representations, the ${\bf 120}$ and  ${\bf 168}$ irrep, were considered  while the weakly attractive ${\bf 4572}$ was disregarded. Moreover, the focus of this previous paper was the study of the recently discovered $\Lambda_b(5912)$ and $\Lambda_b(5920)$ states \cite{Aaij:2012da}. In the present paper we aim at studying the $\bar B N$ and $\bar B \Delta$ interactions to analyze the corresponding scattering amplitudes and, hence, the cross sections for the $\bar B$ propagation in matter. Therefore, we analyze all resonant states appearing in the scattering amplitude stemming from all attractive representations for energies ranging from 5.8 GeV (close to the newly discovered states) up to 6.5 GeV.

We note that to achieve a better description of the $\Lambda_b(5912)$ and $\Lambda_b(5920)$ states reported by the
LHCb Collaboration, we have slightly changed the value of the subtraction point used in the renormalization scheme \cite{GarciaRecio:2012db},
\begin{equation}
\mu^2 =\alpha~(M_{\Sigma_b}^2+m_\pi^2) \  , \label{eq:defalpha}
\end{equation}
with $\alpha=0.967$. We will use the same value  $\alpha$ in all sectors.

We observe that several of the $I=0, J=1/2$ states are very close in energy to the $I=0, J=3/2$ ones. In particular, some of these  $J=1/2$ and $J=3/2$ states form HQSS doublets,  as previously defined. As a formal rule, states with different spin and equal SU(6) and SU(3) labels form a HQSS multiplet \cite{Romanets:2012hm,GarciaRecio:2012db,Garcia-Recio:2013gaa}. We are considering the resonances stemming from  the {\bf 120} and {\bf 168} SU(8) most attractive representations as well as the {\bf 4752} SU(8) irrep, this last one with a much higher multiplicity. Thus, the analysis of the adiabatic evolution of the states from the SU(6)~$\times$~HQSS symmetric point to the physical one 
in order to assign distinct SU(6) and SU(3) labels (in similar way as done in Refs.~\cite{Romanets:2012hm,GarciaRecio:2012db,Garcia-Recio:2013gaa} for the states in the {\bf 120} and {\bf 168}  irreps)  is a much more tedious and difficult task, and is beyond the scope of the present paper. We have, however, restored HQSS in some cases, when the identification was dubious. We find seven HQSS doublets. While the HQSS doublet [$\Lambda_b(5910)$, $\Lambda^*_b(5921)$] was discussed in Ref.~\cite{Romanets:2012hm} and assigned to the newly discovered $J=1/2$ $\Lambda_b(5912)$ and $J=3/2$ $\Lambda^*_b(5920)$ states  \cite{Aaij:2012da},  other HQSS doublets are: [$\Lambda_b(6034)$, $\Lambda^*_b(6045)$],  [$\Lambda_b(6094)$, $\Lambda^*_b(6105)$], [$\Lambda_b(6202)$, $\Lambda^*_b(6207)$],  [$\Lambda_b (6362)$, $\Lambda^*(6373)$], [$\Lambda_b (6464)$, $\Lambda^*(6464)$] and [$\Lambda_b (6516)$, $\Lambda^*(6520)$]. 

Several works have conjectured the existence of bottom baryonic resonances  \cite{Capstick:1986bm,Garcilazo:2007eh,Ebert:2007nw,Karliner:2008sv,Roberts:2007ni,Yamaguchi:2014era}, most of them based on quark models. Recently baryon-meson  calculations in the bottom sector using an extended hidden-gauge model  have been
carried out \cite{Liang:2014eba}. This work considers the interaction of $N\bar B $, $ \Delta\bar B$, $N\bar B^*$ and $ \Delta \bar B^*$ states with their coupled channels. The connection between $\bar B$ and $\bar B^*$ states with nucleon and $\Delta$ baryons is performed by requiring pion exchange, or anomalous terms, which are subleading in the large heavy-quark mass counting. The dynamics of the interaction is, though, different in our approach.  In our model we consider simultaneously all  baryon-pseudoscalar meson ($BP$) and  baryon-vector meson ($BV$) channels, with $J^P=1/2^+$ and $3/2^+$ baryons, using a WT type-interaction that respects SU(6)~$\times$~HQSS symmetry. The potential in the $BP-BP$ and $BV-BV$ sectors in both models is  similar although a larger coupled-channel space  is considered  within our scheme. Moreover, the model of Ref.~\cite{Liang:2014eba} uses a different renormalization scheme and  takes into account a suppression factor in those transitions that involve a $t$-channel 
exchange 
of   a heavy charm vector meson, which is not required from HQSS \cite{GarciaRecio:2012db}.

In Ref.~\cite{Liang:2014eba}, six $\Lambda_b$ and $\Lambda^*_b$ have been found, two of them  associated to the experimental $\Lambda_b(5910)$ and $\Lambda^*_b(5921)$ states with an important coupling to $\bar B^* N$. Moreover, their $\Lambda_b(5821)$, with a strong coupling to $\bar B N$,  was identified with our $\Lambda_b(5798)$ and the $\Lambda_b(5969)$ with a dominant coupling to $\Sigma_b \pi$ was assigned to our  $\Lambda_b(6009)$. However,  this last assignment seems dubious, since our $\Lambda_b(6009)$ does not couple strongly to $\Sigma_b \pi$. The two last states $\Lambda_b(6317)$ and $\Lambda^*_b(6316)$ in Ref.~\cite{Liang:2014eba} couple strongly to $\Sigma_b \rho$. We could assign them to our HQSS doublet  [$\Lambda_b(6202)$, $\Lambda^*_b(6207)$] since we find an important coupling to $\Sigma_b \rho$ and $\Sigma^*_b \rho$, respectively, although our masses are smaller by 100 MeV and the widths are larger. Also, one could assign the $\Lambda_b(6317)$ of Ref.~\cite{Liang:2014eba} to our $\Lambda_
b(6243)$, due to the dominant $\Sigma_b^* \rho$ and $\Sigma_b \rho$ channels. The enlarged coupled-channel space in our model allows for a different composition of the resonant states as compared to the extended hidden-gauge scheme, thus making sometimes difficult a straightforward identification of the states between the two models.

\subsubsection{$\Sigma_b$ and $\Sigma_b^*$ states \rm{($I=1; J=1/2,3/2$)}}

In the $\Sigma_b$ sector, there are 22 channels
\noindent
\begin{center}
\begin{tabular}{c c c c c c c c}
$\Lambda_b \pi$  & $\Sigma_b \pi$    & $N \bar B$  & $N \bar B^*$  & $\Xi_b K $  & $\Sigma_b \eta $   & $\Lambda_b \rho $  &  $\Xi'_b K$ \\
  $\Delta \bar B^*$   & $\Sigma \bar B_s$   &  $\Sigma_b \rho$   & $\Sigma_b \omega$ &     $\Sigma \bar B_s^*$    &  $\Sigma_b^* \rho$    &    $\Sigma_b^* \omega$   &  $\Xi_b K^*$ \\
$\Sigma_b \eta'$    &   $\Sigma^* \bar B_s^*$     &    $\Xi'_b K^*$        &    $\Sigma_b \phi$      
  &   $\Xi_b^* K^*$   &    $\Sigma_b^* \phi$       
\end{tabular}
\end{center}

In the $\Sigma^*_b$ sector, we find 20 channels
\noindent
\begin{center}
\begin{tabular}{cccccccc}
  $\Sigma_b^* \pi$   &   $N \bar B^*$   &   $\Sigma_b^* \eta$    &    $\Lambda_b \rho$    &     $\Xi_b^* K$   &   $\Delta \bar B$   &    $\Delta \bar B^*$     &    $\Sigma_b \rho$ \\
      $\Sigma_b \omega$      &   $\Sigma \bar B_s^*$    &
    $\Sigma_b^* \rho$       &     $\Sigma_b^* \omega$
 &   $\Xi_b K^*$      &    $\Sigma^* \bar B_s$   &        $\Sigma_b^* \eta'$        &     $\Sigma^* \bar B_s^*$    \\
  $\Xi'_b K^*$   &    $\Sigma_b \phi$     &    $\Xi_b^* K^*$     &     $\Sigma_b^* \phi$
\end{tabular}
\end{center}

In both cases the channels are ordered by increasing mass thresholds.

We show in Table \ref{TabPoles1} several $J=1/2$ and $J=3/2$ states ordered by increasing mass, in a similar way as done in Table~\ref{TabPoles0}. As described in the $\Lambda_b$ and $\Lambda^*_b$ sectors,
we can also distinguish several HQSS doublets due to their decay modes and closeness in mass. We find eight HQSS doublets, such as [$\Sigma_b(5812)$, $\Sigma^*_b(5820)$], 
[$\Sigma_b(5971)$, $\Sigma^*_b(5980)$],  [$\Sigma_b(6021)$, $\Sigma^*_b(6028)$], [$\Sigma_b(6217)$, $\Sigma^*_b(6228)$],  [$\Sigma_b(6309)$, $\Sigma^*_b(6319)$],  [$\Sigma_b(6359)$, $\Sigma^*_b(6365)$],
[$\Sigma_b(6469)$, $\Sigma^*_b(6479)$] and  [$\Sigma_b(6512)$, $\Sigma^*_b(6517)$].

It is also interesting to note that our state $\Sigma_b(5904)$ can be interpreted as the counterpart of the $\Sigma^*(1670)$ and $\Sigma_c^*(2549)$  in the strange and charm sectors, respectively. This
assignment is due to the fact that this state has a dominant $\Delta \bar B$ component, in a similar manner as the  $\Sigma^*(1670)$ and $\Sigma_c^*(2549)$ resonances strongly couple to $\Delta \bar K$ and $\Delta D$,
respectively. This state has not been found experimentally yet, but it is a clear case for discovery.

As also mentioned for the $\Lambda_b$ and $\Lambda^*_b$, the straightforward comparison with the predicted $\Sigma_b$ and $\Sigma^*_b$ states of Ref.~\cite{Liang:2014eba} is difficult in some
cases. However, in Ref.~\cite{Liang:2014eba}, the bottom counterpart of the $\Sigma^*(1670)$ and $\Sigma_c^*(2549)$ could be traced to the $\Sigma^*_b(5933)$, 30 MeV above our prediction. Moreover, our
HQSS doublet  [$\Sigma_b(6021)$, $\Sigma^*_b(6028)$] could be identified with $\Sigma_b(6023)$ of Ref.~~\cite{Liang:2014eba}. This state, however, couples most strongly to $\bar B^* \Delta$
while the $\Sigma_b(6021)$ couples dominantly to $\Sigma \bar B_s$ and the $\Sigma^*_b(6028)$ resonance couples strongly to $\Sigma \bar B^*_s$. These two-particle channels are not present in the coupled-channel space of Ref.~\cite{Liang:2014eba},
since states with strangeness and hidden strangeness are not taken into account.

\begin{table*}
\begin{ruledtabular}
\begin{center}
\begin{tabular}{  c  c  c  c  c  }

 $M_{R}$ & $\Gamma_{R}$   & Couplings  & $J$	& Open \\
 (MeV) & (MeV)  &   to main
 channels & & channels \\
 \hline
 
{{5811.8}}  &  0.1  &   $g_{\Sigma_b \pi}=2.$,  ${g_{N \bar B}}=5.7$,
 $g_{N \bar B^*}=3.7$,  $g_{\Delta \bar B^*}=2.8$ &  {{1/2}} & $\Lambda_b \pi$ \\

{{5820.4}} &  0.0  &  $g_{\Sigma_b^* \pi}=1.9$,  $g_{N \bar B^*}=6.9$, 
 ${g_{\Delta \bar B}=1.8}$,  $g_{\Delta \bar B^*}=2.1$ &  {{3/2}} & \\


5903.6  &  0.0  &   ${g_{\Delta \bar B}=6.}$,  $g_{\Delta \bar B^*}=4.5$,  
 $g_{\Sigma_b \omega}=2.7$
 &  3/2 & \\


5909.3 &  0.0  &  $g_{N \bar B^*}=2.3$,  $g_{\Sigma_b^* \eta}=2.2$, 
 $g_{\Delta \bar B^*}=2.3$,  $g_{\Sigma^* \bar B_s}=4.5$,
 $g_{\Sigma^* \bar B_s^*}=5.5$  &  3/2 & \\

 
5911.0 &  0.0  & ${g_{N \bar B}=1.8}$, 
  $g_{\Sigma_b \eta}=2.2$,   $g_{\Xi'_b K}=1.8$,
 $g_{\Sigma \bar B_s}=2.8$,  $g_{\Sigma^* \bar B_s^*}=7.1$ & 1/2 & $\Lambda_b \pi$ \\


5918.4 &  0.0  &   $g_{\Xi_b K}=2.6$,
  $g_{\Sigma \bar B_s}=3.7$,  $g_{\Sigma \bar B_s^*}=6.2$,  $g_{\Sigma^* \bar B_s^*}=1.8$
 &  1/2 & $\Lambda_b \pi$ \\

{{5970.6}}  &  0.5  & $g_{\Xi'_b K}=2.1$,
 $g_{\Sigma \bar B_s}=5.5$,  $g_{\Sigma \bar B_s^*}=3.8$,  
$g_{\Sigma^* \bar B_s^*}=2.7$   & {{1/2}}  & $\Lambda_b \pi$, $\Sigma_b \pi$\\

{{5980.4}}  & 0.4  &    $g_{\Xi_b^* K}=1.9$,  $g_{\Sigma \bar B_s^*}=7.$,
   $g_{\Xi_b K^*}=1.8$,  $g_{\Sigma^* \bar B_s^*}=1.7$ 
  &  {{3/2}} & $\Sigma_b^* \pi$ \\

 6015.0  &  2.6  &   $g_{N \bar B^*}=1.8$,
 $g_{\Sigma_b \rho}=1.7$,   $g_{\Sigma_b^* \rho}=1.9$,  $g_{\Sigma \bar B_s^*}=3.$
 & 1/2 & $\Lambda_b \pi$, $\Sigma_b \pi$\\

{{6020.6}}  &  10.0  &    
 $g_{\Lambda_b \rho}=1.6$,  $g_{\Sigma \bar B_s}=2.2$,  $g_{\Delta \bar B^*}=2.$, $g_{\Xi_b K^*}=1.4$
  &  {{1/2}} & $\Lambda_b \pi$, $\Sigma_b \pi$  \\

{{6027.6}}  &  9.1  &    $g_{\Lambda_b \rho}=1.6$,
 $g_{\Delta \bar B^*}=1.6$,    
 $g_{\Sigma \bar B_s^*}=2.3$,  $g_{\Xi_b K^*}=1.4$&  {{3/2}} & $\Sigma_b^* \pi$ \\

6051.4  &  0.02  &   $g_{\Sigma^* \bar B_s}=5.7$,  $g_{\Xi'_b K^*}=2.7$,
  $g_{\Sigma_b \phi}=1.8$,  $g_{\Sigma^* \bar B_s^*}=5.$ & 3/2 & $\Sigma_b^* \pi$\\

{{6216.7}} & 38.0   &    $g_{\Sigma_b \eta}=1.$,
  $g_{\Sigma_b \omega}=1.5$,  $g_{\Sigma_b^* \omega}=0.8$,  $g_{\Xi_b K^*}=1.1$    & {{1/2}}  & $\Lambda_b \pi$, $\Sigma_b \pi$ \\

{{6227.9}}  &  38.1  &  $g_{\Sigma_b^* \pi}=0.8$, $g_{\Lambda_b \rho}=0.8$,  $g_{\Sigma_b^* \eta}=1.$,  
 $g_{\Sigma_b^* \omega}=1.6$,  $g_{\Xi_b K^*}=1.1$  &  {{3/2}}  & $\Sigma_b^* \pi$ \\

6256.0  &  41.4  &   $g_{\Lambda_b \pi}=0.7$,   
 $g_{\Sigma_b \omega}=1.5$,  $g_{\Sigma_b^* \omega}=2.2$   &  1/2 & $\Lambda_b \pi$, $\Sigma_b \pi$, $N \bar B$ \\

{{6308.8}}  & 6.6  &    $g_{\Sigma_b \eta}=0.9$,
$g_{\Sigma_b \omega}=1.6$,  $g_{\Sigma_b^* \omega}=1.3$,  $g_{\Xi_b K^*}=1.$ &  ${{1/2}}$  & $\Lambda_b \pi$, $\Sigma_b \pi$, $N \bar B$ \\
& & & & $N \bar B^*$, $\Xi_b K$\\

{{6319.0}} & 6.1 &    $g_{\Sigma_b^* \eta}=1.$,
 $g_{\Sigma_b^* \omega}=1.8$,  $g_{\Xi_b K^*}=1.$  &  {{3/2}} & $\Sigma_b^* \pi$, $N \bar B^*$ \\

{{6359.3}}  &  0.2  &     $g_{\Lambda_b \rho}=1.$, 
 $g_{\Sigma_b \rho}=1.7$, $g_{\Sigma_b^* \rho}=1.2$
&  {{1/2}}  & $\Lambda_b \pi$, $\Sigma_b \pi$, $N \bar B$, $N \bar B^*$\\

{{6364.7}}  &  0.1  &  $g_{\Lambda_b \rho}=1.$,
$g_{\Sigma_b \rho}=1.1$,  $g_{\Sigma_b^* \rho}=1.8$   & {{ 3/2}} & $\Sigma_b^* \pi$, $N \bar B$\\

6401.7   &   28.7  &          
 $g_{\Sigma \bar B_s^*}=1.$,  $g_{\Sigma_b \phi}=1.3$,   $g_{\Sigma_b^* \phi}=1.8$,  $g_{\Xi_b^* K^*}=1.2$
   &  1/2 & $\Lambda_b \pi$, $\Sigma_b \pi$, $N \bar B$, \\
  & &  & & $N \bar B^*$, $\Xi_b K$, $\Sigma_b \eta$, $\Lambda_b \rho$ \\

6409.8  &  0.1  &   
  $g_{\Sigma_b \rho}=1.9$,  $g_{\Sigma_b \omega}=1.2$,
 $g_{\Sigma_b^* \rho}=1.1$  & 3/2 & $\Sigma_b^* \pi$, $N \bar B^*$, $\Lambda_b \rho$, $\Sigma_b^* \eta$\\

{{6469.2}}  &  16.4   &    
$g_{\Xi'_b K^*}=1.9$,  $g_{\Sigma_b \phi}=1.5$,  
$g_{\Xi_b^* K^*}=1.1$ & {{1/2}} & $\Lambda_b \pi$, $\Sigma_b \pi$, $N \bar B$, $N \bar B^*$\\
 &  &      & & $\Xi_b K$, $\Sigma_b \eta$, $\Lambda_b \rho$,  $\Xi'_b K$ \\

{{6478.6}}  &  15.1  &     
   $g_{\Xi'_b K^*}=0.9$, $g_{\Sigma_b^* \phi}=1.5$,  $g_{\Xi_b^* K^*}=2.$  & {{3/2}}& $\Sigma_b^* \pi$,  $N \bar B^*$, $\Lambda_b \rho$\\
 &  &    & & $\Sigma_b^* \eta$,  $\Delta \bar B$, $\Xi_b^* K$ \\

 {{6512.5}}  &  0.6   &  
     $g_{\Xi_b K^*}=1.4$,    $g_{\Xi'_b K^*}=1.$,   $g_{\Sigma_b \phi}=1.7$, $g_{\Sigma_b^* \phi}=1.2$&  ${{1/2}}$  & $\Lambda_b \pi$, $\Sigma_b \pi$, $N \bar B$  \\
 & &          & & $N \bar B^*$, $\Xi_b K$,  $\Sigma_b \eta$, $\Lambda_b \rho$ \\

{{6517.0}}  &  0.6  &    
 $g_{\Xi_b K^*}=1.4$,  $g_{\Sigma_b \phi}=1.2$,  $g_{\Sigma_b^* \phi}=1.8$,  $g_{\Xi_b^* K^*}=1.1$  &  {{3/2}} &  $\Sigma_b^* \pi$, $N \bar B^*$, $\Lambda_b \rho$ \\
 &  &    & & $\Sigma_b^* \eta$, $\Delta \bar B$, $\Xi_b^* K$ \\

{{6542.2}}  & 1.4   &    $g_{\Xi'_b K^*}=1.$,  $g_{\Sigma_b \phi}=0.9$,
  $g_{\Sigma_b^* \phi}=1.6$,  $g_{\Xi_b^* K^*}=1.7$    &   {{1/2}}   & $\Lambda_b \pi$, $\Sigma_b \pi$, $N \bar B$, $N \bar B^*$ \\
 &  &  & & $\Xi_b K$, $\Sigma_b \eta$, $\Lambda_b \rho$, $\Xi'_b K$ \\

{{6549.0}}  &  0.02  &  $g_{\Xi'_b K^*}=1.4$,  $g_{\Sigma_b \phi}=2.$,  $g_{\Sigma_b^* \phi}=1.3$  & {{3/2}} & $\Sigma_b^* \pi$, $N \bar B^*$, $\Lambda_b \rho$ \\
 &  &    &  & $\Sigma_b^* \eta$, $\Delta \bar B$, $\Xi_b^* K$, $\Delta \bar B^*$\\



\end{tabular}
\end{center}
\end{ruledtabular}
\caption{As in Table~\ref{TabPoles0}, but for  $\Sigma_b$ and $\Sigma_b^*$ baryon resonances ($I=1; ~J=1/2, ~J=3/2$).
}
\label{TabPoles1}
\end{table*}

\subsubsection{$(I=2; J=3/2)$ states}

In this sector we have 5 channels
\noindent
\begin{center}
\begin{tabular}{ccccc}
 $\Sigma_b^* \pi$  & $\Delta \bar B$ & $\Delta \bar B^*$ & $\Sigma_b \rho$ & $\Sigma_b^* \rho$
\end{tabular}
\end{center}
again ordered by increasing mass thresholds.

\begin{table*}
\begin{ruledtabular}
\begin{center}
\begin{tabular}{  c  c  c  c  c  }

 $M_{R}$ & $\Gamma_{R}$   & Couplings  & $J$	& Open \\
 (MeV) & (MeV)  &   to main
 channels & & channels \\ 
\hline

5907.8   &  0.0  &   $g_{\Sigma_b^* \pi}=2.2$,  ${g_{\Delta \bar B}=4.8}$,    $g_{\Delta \bar B^*}=5.9$  &  3/2 &  \\

6048.6  &  0.05  &    ${ g_{\Delta \bar B}=5.4}$,    $g_{\Delta \bar B^*}=4.6$,
$g_{\Sigma_b \rho}=2.8$,   $g_{\Sigma_b^* \rho}=1.3$  &  3/2  & $\Sigma_b^* \pi$ \\

6395.1  & 31.1   &
$g_{\Sigma_b \rho}=1.1$,   $g_{\Sigma_b^* \rho}=2.5$ 
&   $3/2$ & $\Sigma_b^* \pi$ \\

\end{tabular}
\end{center}
\end{ruledtabular}
\caption{As in Table~\ref{TabPoles0}, but for baryon resonances in the $I=2$; $J=3/2$ sector.}
\label{TabPolesBbarDel2}
\end{table*}

As seen in Table \ref{TabPolesBbarDel2}, we obtain three resonances with masses 5909  MeV, 6049 MeV and  6395 MeV. Whereas the first two states are mainly formed by $\Delta \bar B$ and
$\Delta \bar B^*$, the last one mostly couples to $\Sigma_b^* \rho$. In Ref.~\cite{Liang:2014eba} it was indicated that the $I=2$ sector is repulsive and, thus, no dynamically-generated
states can be found.  However, in this previous work  the $\Sigma_b \rho$ and $\Sigma_b^* \rho$ channels were not considered in the coupled basis. In our model, the inclusion of these two channels
provides some attraction and allows for the formation of the three dynamically-generated baryon-meson states.


\section{$\bar B$ meson propagation in hadronic matter}
 \label{sec:transport}

In the last section we have obtained a realistic description of the interactions of a $\bar B$ meson
with light mesons and baryons, by means of unitarized effective field theories.
An immediate application is to study the $\bar B$ meson propagation in a dense and hot medium composed of lighter mesons and baryons. 

If a hadron mixture --such as a hadron gas in heavy-ion collisions--  is out
of equilibrium, the scattering of a heavy meson with other species implies momentum loss as well as
entropy production. These effects are encoded into the transport coefficients of heavy mesons and, in particular, in the medium drag force and the diffusion coefficients~\cite{landau1981course}.

In a collective description, the distribution function of $\bar B$ mesons, $f(t,{\bf p})$,
obeys a Boltzmann-like transport equation, provided that the system is dilute enough and there are no
correlations between collisions.  In this picture, the heavy meson behaves as a Brownian particle suffering from collisions with the bath's particles. In the limit of a large Brownian mass (in comparison to the other masses), 
the transport equation can be recasted into a  Fokker-Planck equation:
\be \frac{\pa f(t,{\bf p})}{\pa t} = \frac{\pa}{\pa p_i} \left\{ F_i (\mathbf{p}) f(t,{\bf p}) + \frac{\pa}{\pa p_j} \left[ \Gamma_{ij} (\mathbf{p}) f(t,\bf{p}) \right] \right\} \ ,\ee
with $i,j=1,2,3$ the spatial indices. The quantity $F_i$ is the drag force, which is a function of the heavy-meson momentum,
\be F_i (\mathbf{p})= \int d\mathbf{k} \ w(\mathbf{p},\mathbf{k}) \ k_i \ , \ee
and $\Gamma_{ij}$ is the momentum diffusion matrix~\cite{landau1981course,Abreu:2011ic},
\be \Gamma_{ij} (\mathbf{p})= \frac{1}{2} \int d\mathbf{k} \ w(\mathbf{p},\mathbf{k}) \ k_i k_j \ . \ee
The collision rate $w(\mathbf{p},\mathbf{k})$ is a remnant of the collision integral in the Boltzmann equation.
It reads
\begin{widetext}
\ba w(\mathbf{p},\mathbf{k}) & =&  g_l \int \frac{d^3q}{(2\pi)^9} \ n_{F,B} (E_l(q),T) 
 \left[ 1\pm n_{F,B} (E_l(q+k),T) \right]   \frac{1}{2E_{\bar B}(p)} \frac{1}{2E_l(q)} 
\frac{1}{2E_{\bar B}(p-k)} \frac{1}{2E_l(q+k)} \nn \\
&\times& (2\pi)^4 \delta (E_{\bar B}(p)+E_l(q)-E_{\bar B}(p-k)-E_l(q+k)) 
 \overline{|\mathcal{M}^2|} \ , 
\label{omega}
\ea
\end{widetext}
where $\bar B$ represents the bottom meson and $l$ the light hadron of the thermal bath. The quantity
$g_l$ stands for the spin-isospin degeneracy factor of the light hadron and $n_{F,B} (E_l,T)$ is 
the light hadron equilibrium distribution function that follows Fermi-Dirac or Bose-Einstein statistics.  The invariant
scattering matrix element is given by $\mathcal{M}$. This is computed as  
\ba
\mathcal{M}_{ij} (\sqrt{s})& = & \gamma_i^{1/2} \gamma_j^{1/2} \ T_{ij} (\sqrt{s}) \ ,
\ea
once the scattering amplitude $T_{ij}$ of Eq.~(\ref{bs}) is known, with $\gamma_i=1$ for meson-meson scattering and
$\gamma_i= 2 M_i$, with $M_i$  the mass of the baryon, for baryon-meson scattering.

Assuming an isotropic bath, the transport coefficients $F_i (\mathbf{p})$ and $ \Gamma_{ij} (\mathbf{p})$ can be written as
\ba F_i (\mathbf{p}) & = & F(p) \  p_i \ , \\
\Gamma_{ij} (\mathbf{p}) &= & \Gamma_0 (p) \left( \delta_{ij} - \frac{p_i p_j}{p^2}\right) + \Gamma_1 (p) \ \frac{p_i p_j}{p^2} \ , 
\ea
in terms of three scalar functions, $F(p)$, $\Gamma_0(p)$ and $\Gamma_1(p)$, given by
\ba \label{eq:Fcoeff}
F(p) & =& \int d\mathbf{k} \ w(\mathbf{p},\mathbf{k}) \frac{k_ip^i}{p^2} \ , \\
\label{eq:G0coeff} \Gamma_0(p) & =& \frac{1}{4} \int d\mathbf{k} \ w(\mathbf{p},\mathbf{k}) \left[ \mathbf{k}^2
- \frac{(k_ip^i)^2}{p^2} \right] \ ,  \\
\label{eq:G1coeff} \Gamma_1(p) & =& \frac{1}{2} \int d\mathbf{k} \ w(\mathbf{p},\mathbf{k}) \frac{(k_ip^i)^2}{p^2} \ . 
\ea

In the so-called {\it static limit} (where the $\bar B$-meson momentum goes to zero) only one of the three coefficients is
independent. On one hand, the two diffusion coefficients become degenerate, 
\be \label{eq:staticdiffusion} \lim_{p\rightarrow 0} \left[ \Gamma_0(p) - \Gamma_1 (p) \right] = 0 \ . \ee
On the other hand, the Einstein relation relates $F$ with $\Gamma=\Gamma_0=\Gamma_1$ as
\be \label{eq:einstein} \lim_{p\rightarrow 0} F(p) = \frac{\Gamma(p)}{m_BT} \ . \ee
Apart from the transport coefficients, there exist other quantities of interest. The relaxation time $\tau_R$ is defined
as the inverse of the drag force:
\be \label{eq:taur} \tau_R = \frac{1}{F} \ , \ee
and corresponds to the characteristic time of relaxation for the momentum
 distribution~\cite{Torres-Rincon:2013nfa}. Moreover, the spatial diffusion coefficient (defined in the static limit)
\be \label{eq:dx} D_x= \lim_{p \rightarrow 0} \frac{\Gamma (p)}{m_B^2 F^2 (p)} \ , \ee
measures the homogenization speed of bottom mesons in the position space~\cite{Torres-Rincon:2013nfa} (q.v. \cite{landau1981course} on diffusion phenomena).

\subsection{Transport coefficients of $\bar B$ mesons}

In this section we present our results on the transport coefficients for a $\bar B$ meson. 
As opposed to the charmed analogue in Ref.~\cite{Tolos:2013kva} we
restrict ourselves to the case of nearly vanishing baryochemical potential. This is due to the fact that the large mass of $\bar B$ mesons makes unlikely their production in colliders such
as FAIR or NICA \cite{nica}, where the finite-$\mu_B$ part of the Quantum Chromodynamics (QCD) phase diagram will be explored.
Only for high-energy colliders such as RHIC or LHC, there is enough initial energy to produce $\bar B$ mesons.

The magnitude of the coefficients in Eqs.~(\ref{eq:Fcoeff},\ref{eq:G0coeff},\ref{eq:G1coeff}) is roughly
determined by the product of the density of light particles and the collision rate (see also
Eq.~(\ref{eq:nonre}) for a nonrelativistic estimate in kinetic theory). Therefore, the cross
section represents a fundamental piece in the computation of the drag force and diffusion 
coefficients. Thus, the presence of resonant structures, as those described in Secs.~\ref{sec:mes} and \ref{sec:bar}, will strongly affect the final values of the transport coefficients.

\begin{figure}
\centering
\includegraphics[width=0.45\textwidth]{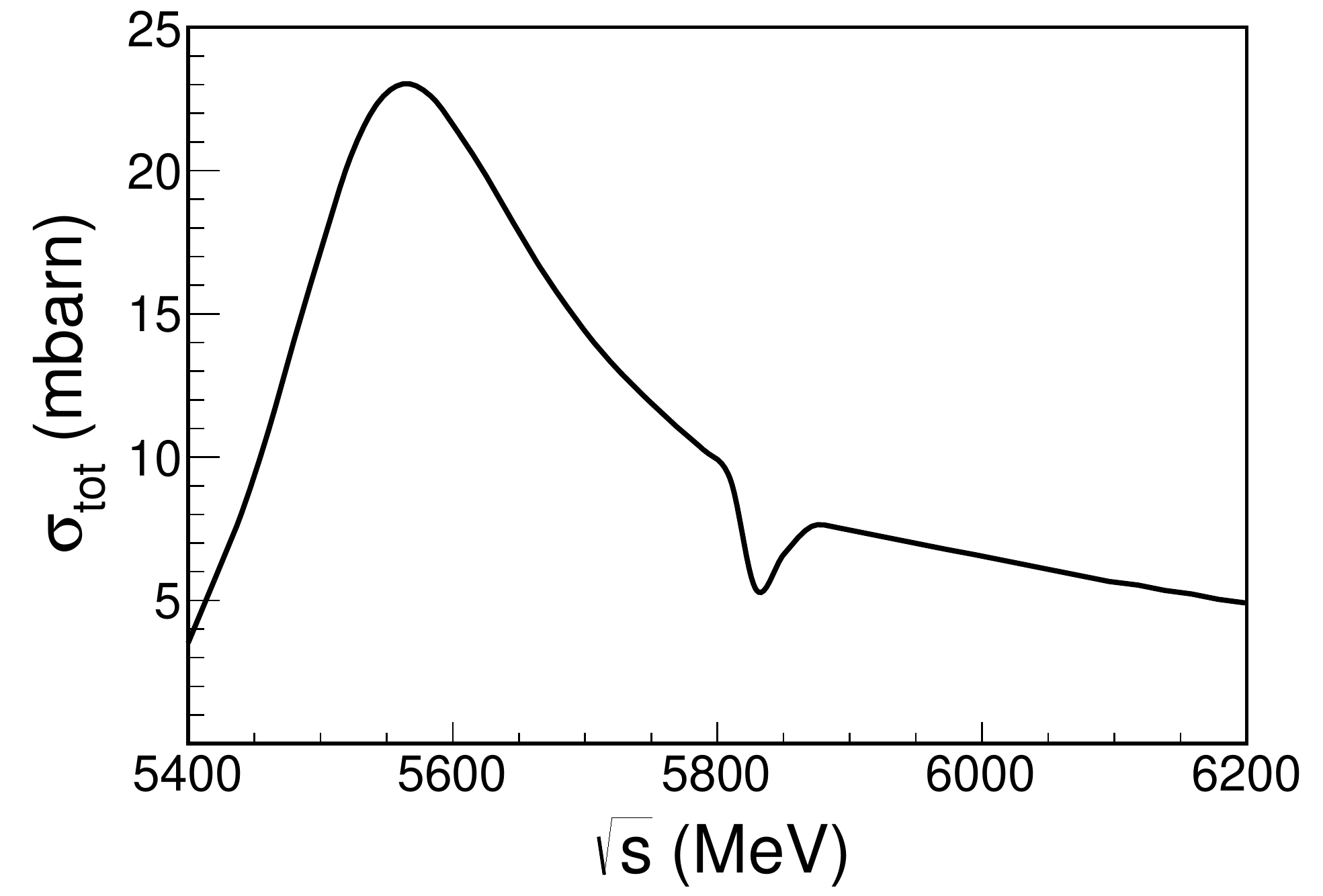}
\includegraphics[width=0.45\textwidth]{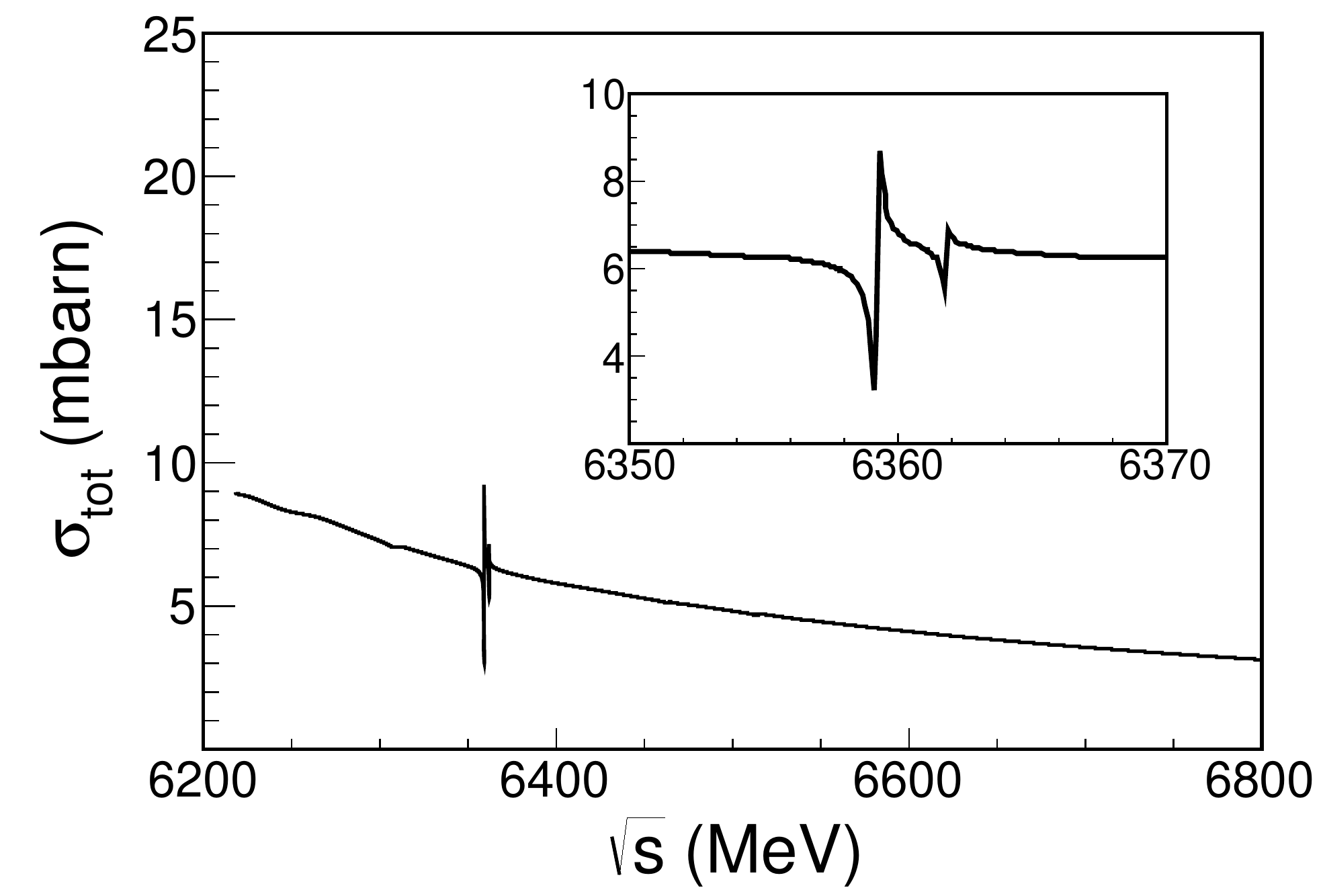}
\caption{\label{fig:av_cs_bpi}
Upper panel: ${\bar B}\pi$ isospin averaged total cross section as a function of the C.M. energy. Lower panel: same for the 
$N{\bar B}$ scattering.}
\end{figure}

In Fig.~\ref{fig:av_cs_bpi} we plot the $\bar B \pi$ and $N \bar B $ elastic cross sections.  These are the dominant cross sections due to a major abundance of pions (nucleons) with respect to the other species in the meson (baryon) sectors. 

In the upper panel, we show the isospin averaged elastic cross section for ${\bar B}\pi$ scattering. This cross section can be compared to our previous computation
in Ref.~\cite{Abreu:2012et}.  As seen in~\cite{Abreu:2012et}, we find that the cross section is  dominated by the presence of the $B_0$ resonance at $\sqrt{s}=5530$ MeV.
There exist, however, two clear differences between our current calculation and the previous one. First, the overall
magnitude of the current cross section is slightly smaller than that in our previous reference.  This is related to new choice of low-energy constants $h_3, h_5$ and the
subtraction constant. The second difference is the clear depression around $\sqrt{s}=5830$ MeV. This feature corresponds to the opening of the  ${\bar B}_s \bar{K}$ channel at $\sqrt{s}=5862$ MeV as the resonance at 5827 MeV starts to form. 

In the lower panel  we display the isospin-averaged $N {\bar B}$ cross section. Note that the magnitude of the cross section is of the same order as the one for the pions away from the $B_0$ peak.
The behavior of the cross section is quite smooth except in the energy region around $\sqrt{s}=6360$ MeV, where the $I=0$ 6361 MeV and the $I=1$ 6359 MeV resonances appear. These states
couple to $N \bar B$ having a very small width and, thus, the average cross section varies abruptly in a small energy domain.

The cross sections in Fig.~\ref{fig:av_cs_bpi} (to be precise, the squared scattering amplitudes) together with those for the $\bar B$ mesons interacting with $K$, $\bar K$, $\eta$ and $\Delta$ are needed to calculate the transport coefficients of Eqs.~(\ref{eq:Fcoeff},\ref{eq:G0coeff},\ref{eq:G1coeff}). We start by considering the zero baryochemical potential, $\mu_B=0$ case, in order to compare with our
previous work in Ref.~\cite{Abreu:2012et}~\footnote{The scattering amplitudes and, hence, cross sections might be modified in matter due to
finite density and temperature effects. The study of these modifications is, however, left for future work in order to carry a detailed
many-body calculation of the scattering amplitudes in matter. This effect is, in any case, subdominant.}.

\begin{figure}
\centering
\includegraphics[width=0.45\textwidth]{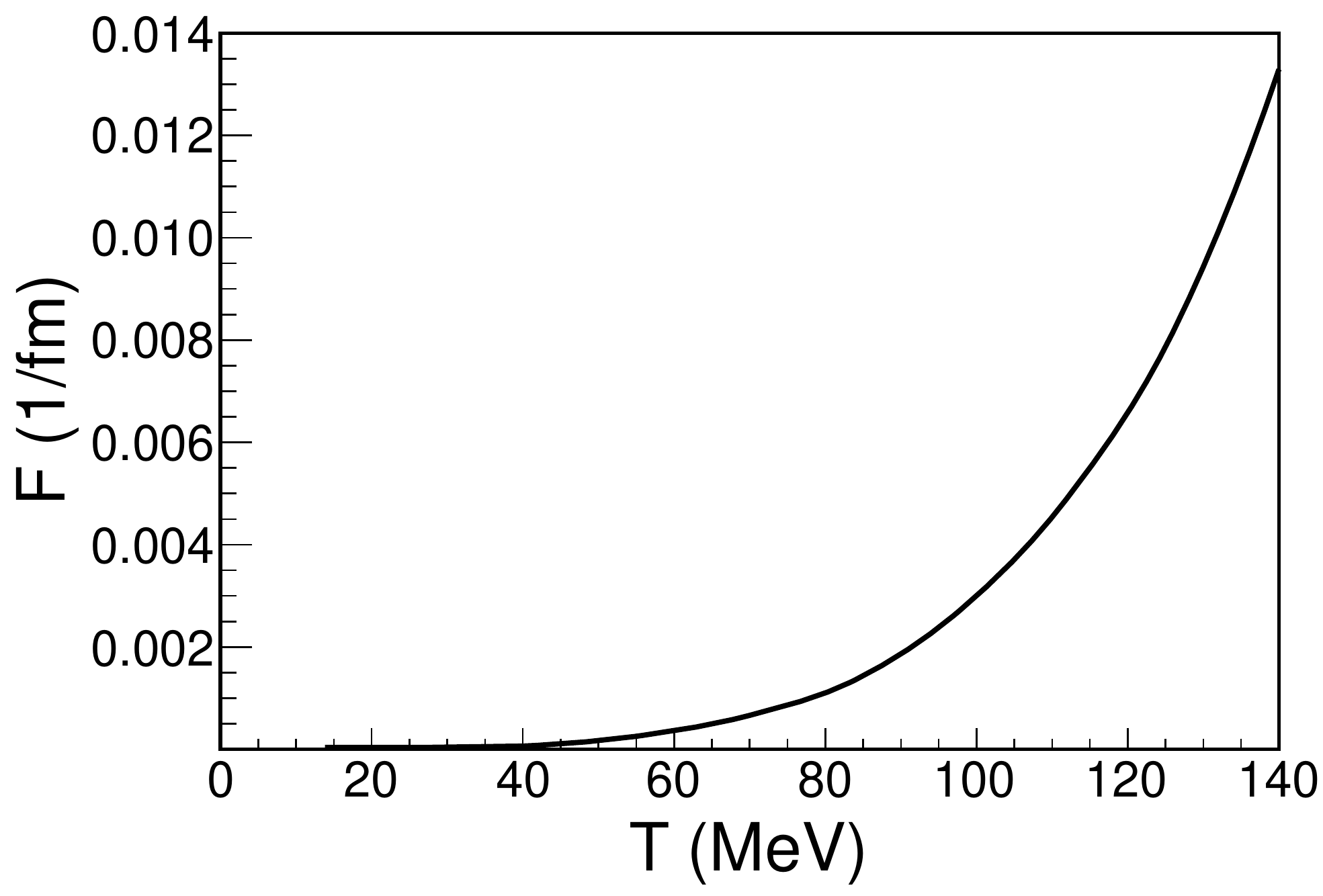}
\includegraphics[width=0.45\textwidth]{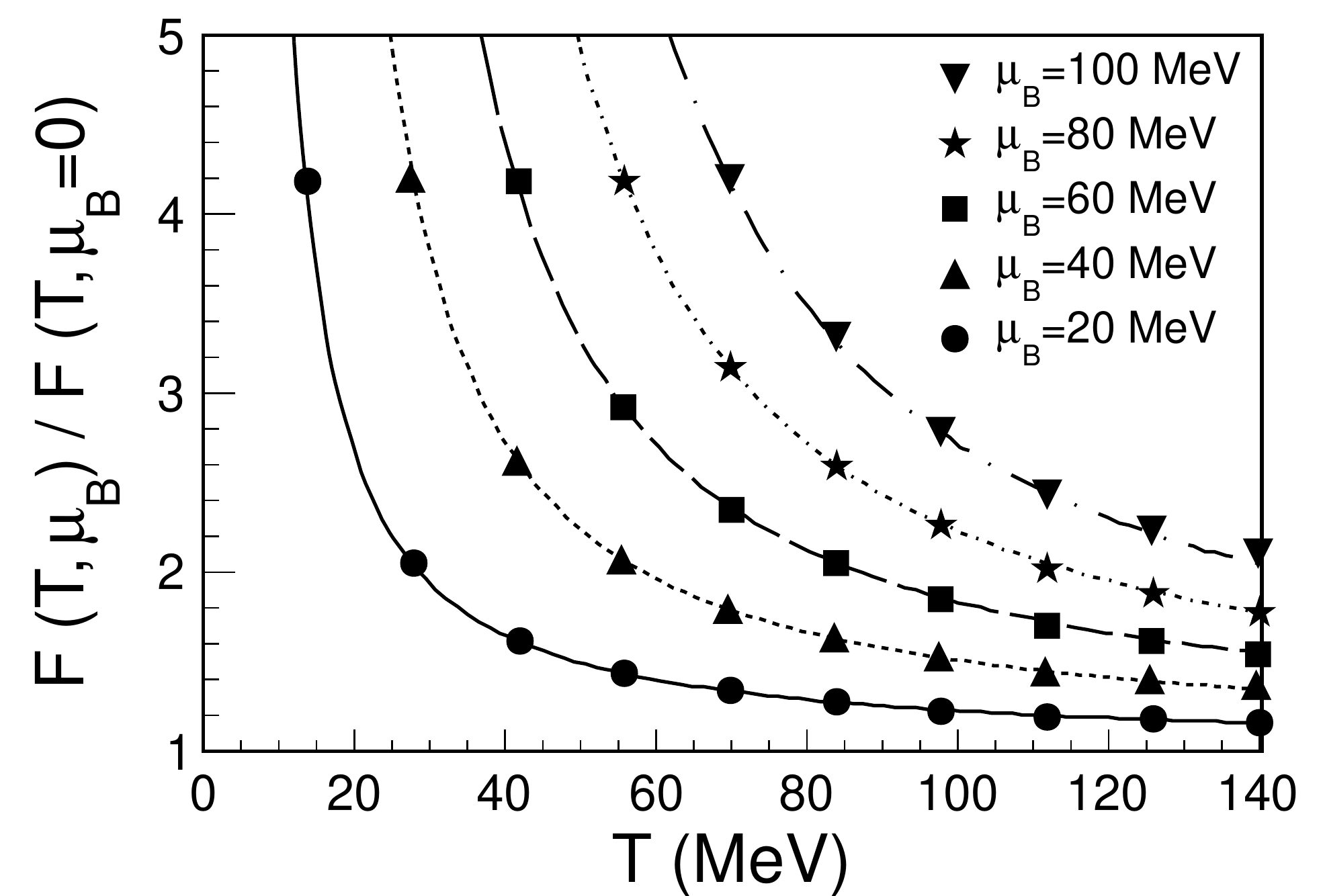}
\caption{\label{fig:Fmu0}
Upper panel: Drag force of a ${\bar B}$ meson in a bath of $\pi,K,\bar{K}$ and $\eta$ mesons.
The baryon chemical potential is set to $\mu_B=0$ and the momentum of the heavy meson to $p=100$ MeV.
Lower panel: Drag force coefficient as a function of temperature for several $\mu_B$ (normalized
to the $\mu_B=0$ case).}
\end{figure}

In the upper panel of Fig.~\ref{fig:Fmu0} we show the drag force $F$ as a function of temperature ($T$) at $\mu_B=0$ and $p=100$ MeV.
Only the $\bar B$-light meson ($\pi,K,{\bar K},\eta$) scattering has been taken into account in this case.

In spite of the addition of more channels in the unitarization procedure and the modification of
the LECs and the subtraction point, the final result of the drag force is similar to the one
in Ref.~\cite{Abreu:2012et}. As seen in  \cite{Abreu:2012et}, the calculations of \cite{Das:2011vba} differ from our results due to the simplified input used for the scattering amplitudes.  We have also checked that the inclusion of the $\bar B$ scattering with $N$ and $\Delta$ makes no
appreciable difference in the $\mu_B=0$ case. As a matter of fact, the resulting curve lies on the top 
of the one presented here, making the baryonic contribution totally negligible. With regard to the diffusion coefficients in the static limit ($\Gamma_0=\Gamma_1$),
the Einstein relation provides them in terms of the drag force $F$.

In order to analyze the contribution of baryons to the transport coefficients, one has to increase the baryochemical potential. However, a large 
baryochemical potential is of very limited interest as $\bar B$ mesons are difficult to 
produce in low-energy heavy-ion collisions, where the finite $\mu_B$ regime  of the QCD phase diagram is probed.

In the lower panel of Fig.~\ref{fig:Fmu0} we plot the drag force at different baryochemical
potentials for the same range of temperatures as the upper one and again at fixed $p=100$ MeV. As the absolute value of
this coefficient is negligible with respect to the contribution from mesons, we have decided to
normalize it with respect to the $\mu_B=0$ case. In this way, we can verify the simple relation,
\be \label{eq:dependence} F(T,\mu_B) =z(\mu_B) \ F(T,\mu_B=0) \ , \ee
with $z=e^{\mu_B/T}$ being the fugacity~\cite{Tolos:2013kva}. The numerical results using
Eq.~(\ref{eq:Fcoeff}) are shown with symbols for $\mu_B=20,40,60,80,100$ MeV. On top of the computation, we have plotted the analytical function $e^{\mu_B/T}$ for the same values of the baryochemical potential. The agreement between the two
is excellent, providing a numerical check of Eq.~(\ref{eq:dependence}) (a similar expression
also holds for the diffusion coefficient). Note that this expression is only valid for the pure baryonic contribution, still very
small compared with the mesonic contribution for low temperatures (Boltzmann suppression) and low baryochemical potentials (small net baryonic density).

\begin{figure}
\centering
\includegraphics[width=0.45\textwidth]{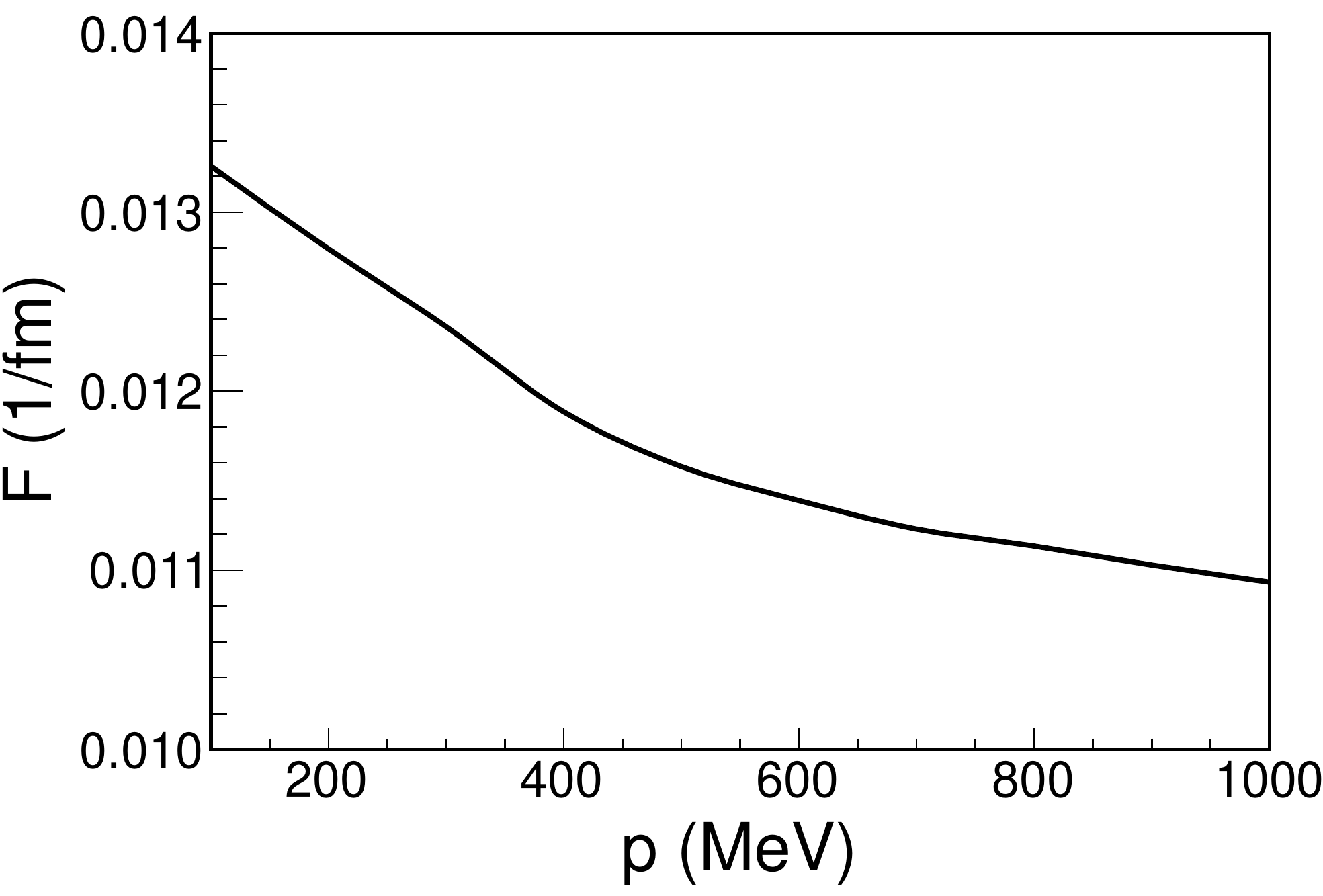}
\includegraphics[width=0.45\textwidth]{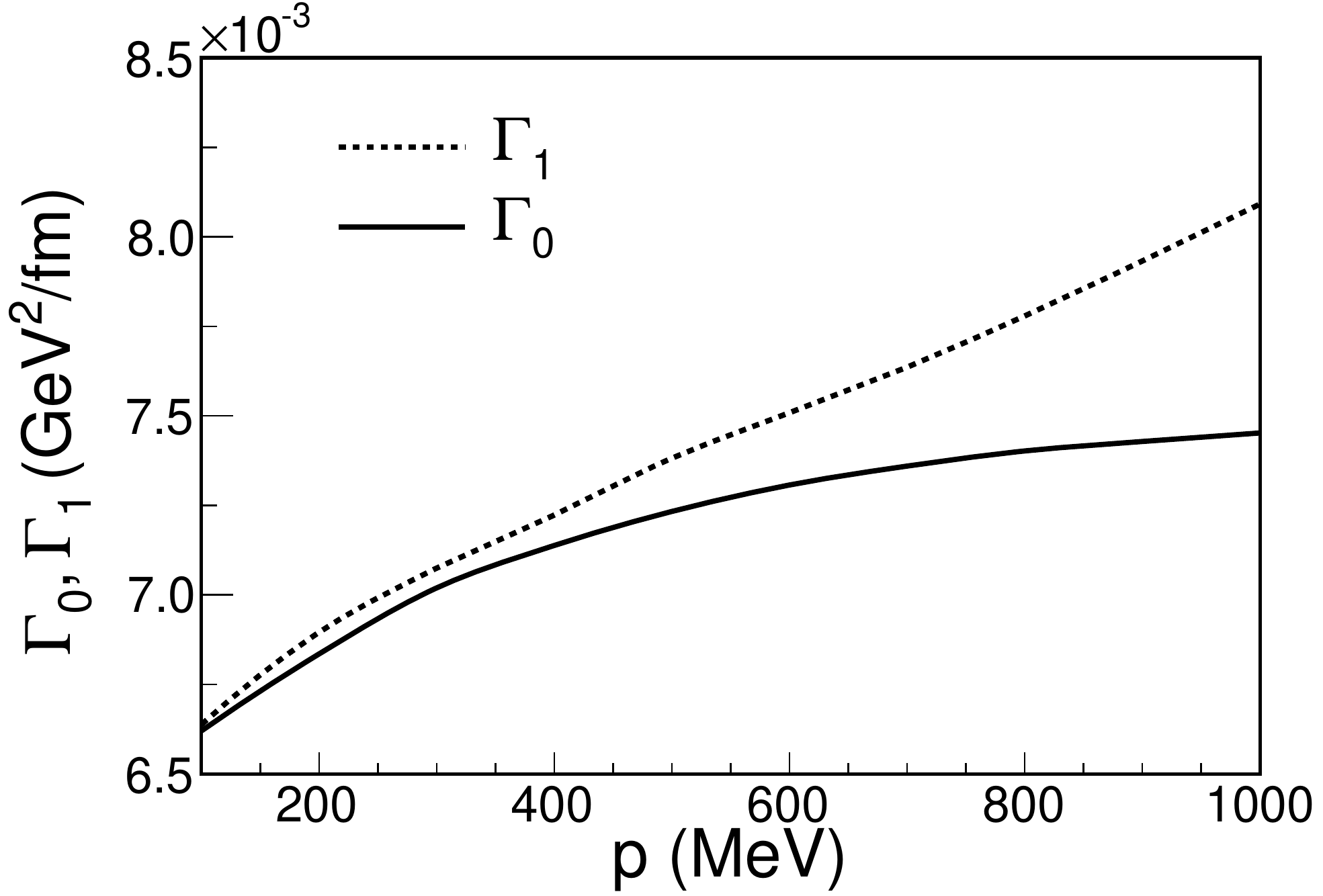}
\caption{\label{fig:coeffs_p}
Drag force and diffusion coefficients of a ${\bar B}$ meson in a bath composed of $\pi,K,\bar{K},\eta$ mesons and $N$ and $\Delta$ baryons.
The baryon chemical potential is fixed at $\mu_B=0$ and the temperature at $T=140$ MeV. The heavy meson momentum
runs from the static limit (taken at $p=100$ MeV) to $p=1$ GeV.}
\end{figure}

In what follows we will present our results of the transport coefficients including all species ($\pi,K,\bar{K},\eta,N,\Delta$). First, we explore their momentum dependence at constant
$\mu_B=0$ and $T=140$ MeV. The results for the three transport coefficients
are shown in Fig.~\ref{fig:coeffs_p}. In this case, the two diffusion coefficients are not
degenerate anymore, although the fluctuation-dissipation theorem still relates the three transport coefficients~\cite{landau1981course,Abreu:2011ic}.
These results are compatible with the ones in Ref.~\cite{Abreu:2012et}, being our result systematically
smaller because the temperature is now $T=140$ MeV. 

So far we have shown the transport coefficients that appear in the Fokker-Planck equation. These coefficients serve as inputs for the numerical propagation of the heavy meson in a hadronic environment based on Langevin dynamics~\cite{Lang:2012cx,Lang:2013cca}.
Alternatively, we present other quantities that possess a more physical insight. In particular,
we pay attention to the relaxation time $\tau_R$ and the spatial diffusion coefficient $D_x$.
These coefficients can be computed in terms of $F$, $\Gamma_0$ and $\Gamma_1$, according to Eqs.~(\ref{eq:taur},\ref{eq:dx}). 
We concentrate on physical trajectories in the QCD phase diagram for the hadronic medium
created at RHIC/LHC collisions, with a large entropy per baryon being constant. As a limiting case, we
present results for a typical FAIR trajectory at its highest energy, with a fixed entropy per baryon
around $s/n_{B}=30$~\cite{Bravina:2008ra,Ejiri:2005uv}.

\begin{figure}
\centering
\includegraphics[width=0.45\textwidth]{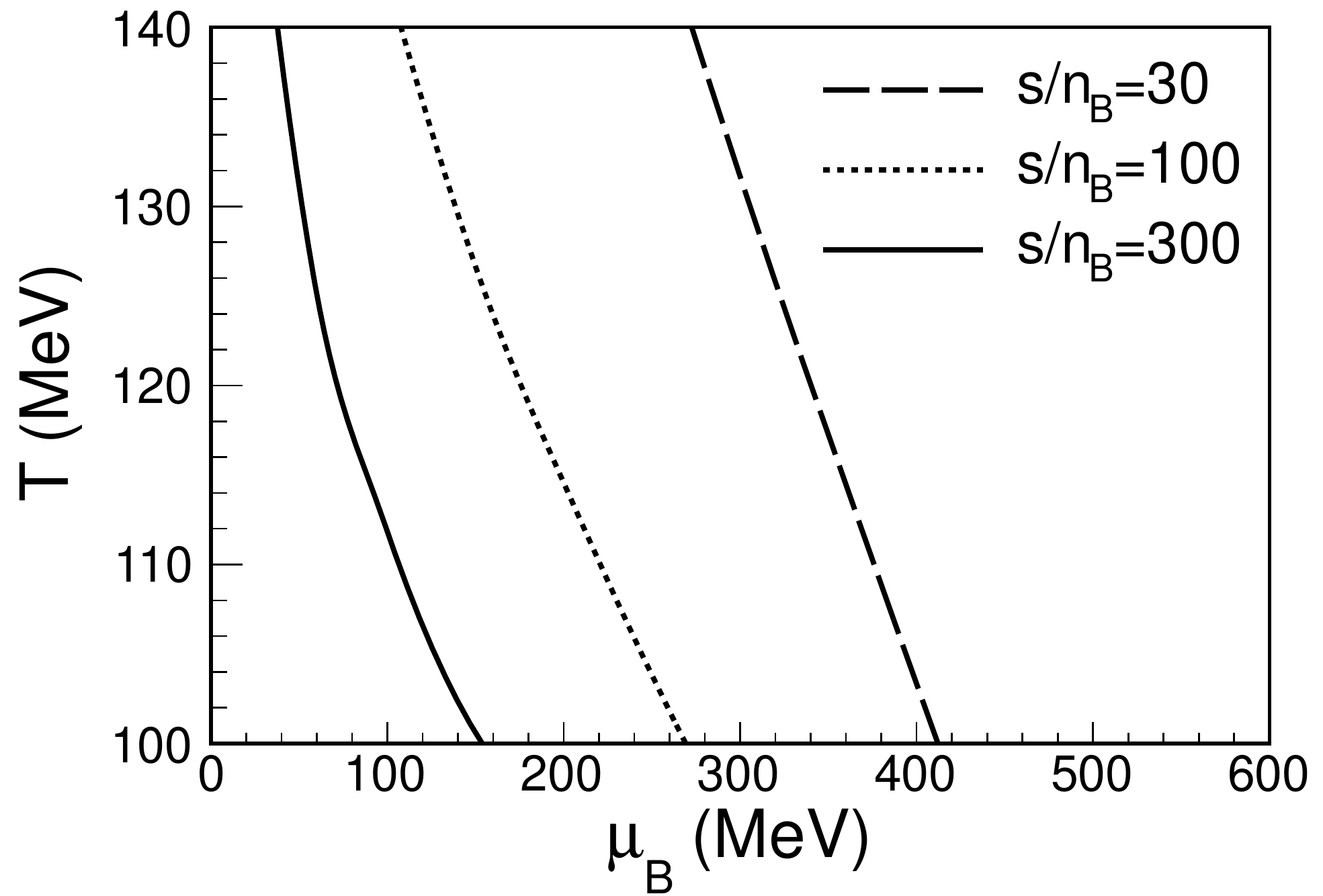}
\caption{\label{fig:isen}
Upper panel: Isentropic trajectories on the phase diagram for collisions at RHIC ($s/n_B=100,300$) and
FAIR at its highest expected energy $s/n_B=30$.}
\end{figure}

Three characteristic trajectories are shown in Fig.~\ref{fig:isen} for fixed entropy
per baryon $s/n_{B}=30, 100$ and $300$. The last two are the predicted values for collisions at
RHIC~\cite{Ejiri:2005uv}. At high $T$, the lines get closer to the $\mu_B=0$ trajectory (thermal evolution of the early
universe) as long as we increase the entropy per baryon. At low temperatures all the curves bend
towards large $\mu_B$.

\begin{figure}
\centering
\includegraphics[width=0.45\textwidth]{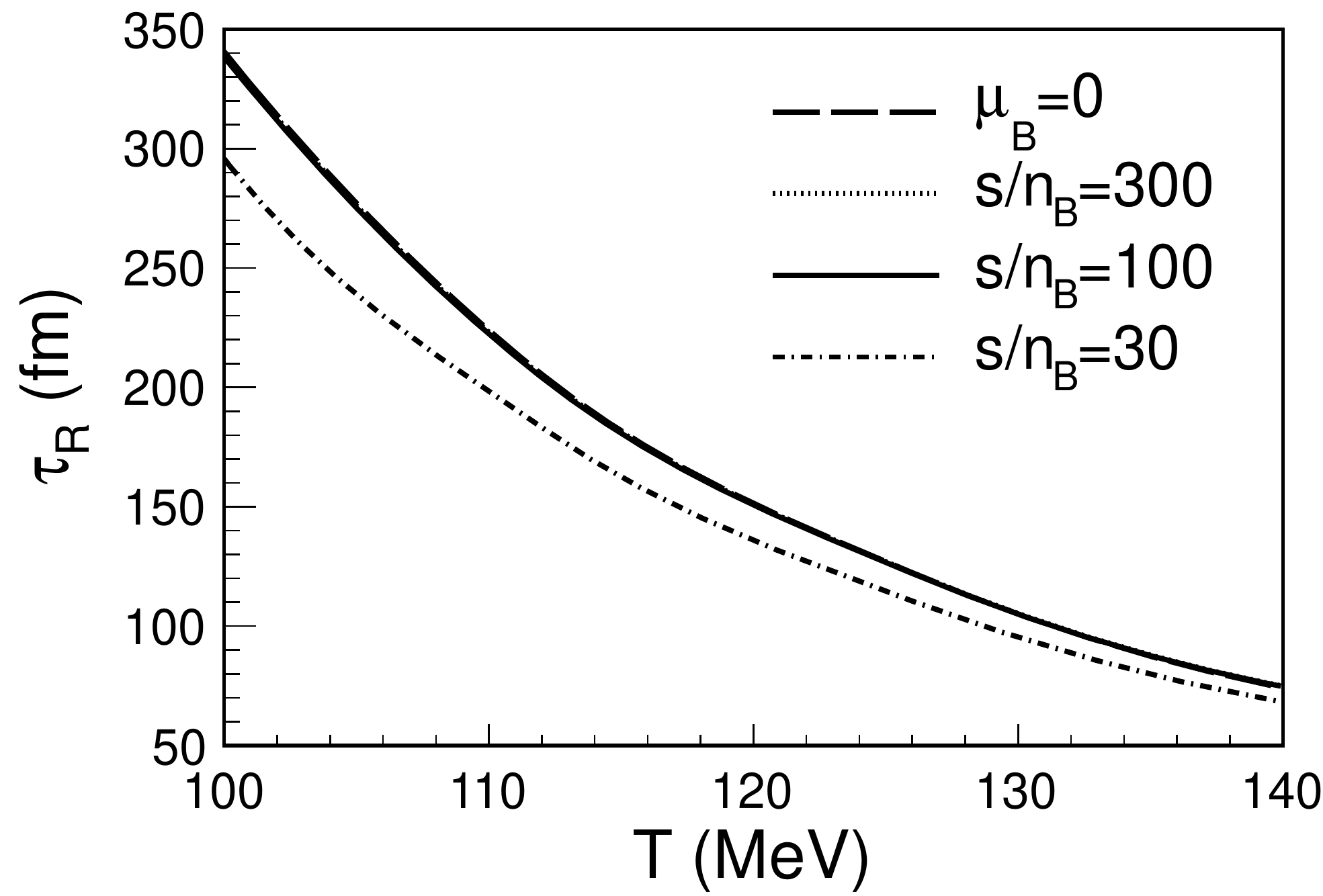}
\includegraphics[width=0.45\textwidth]{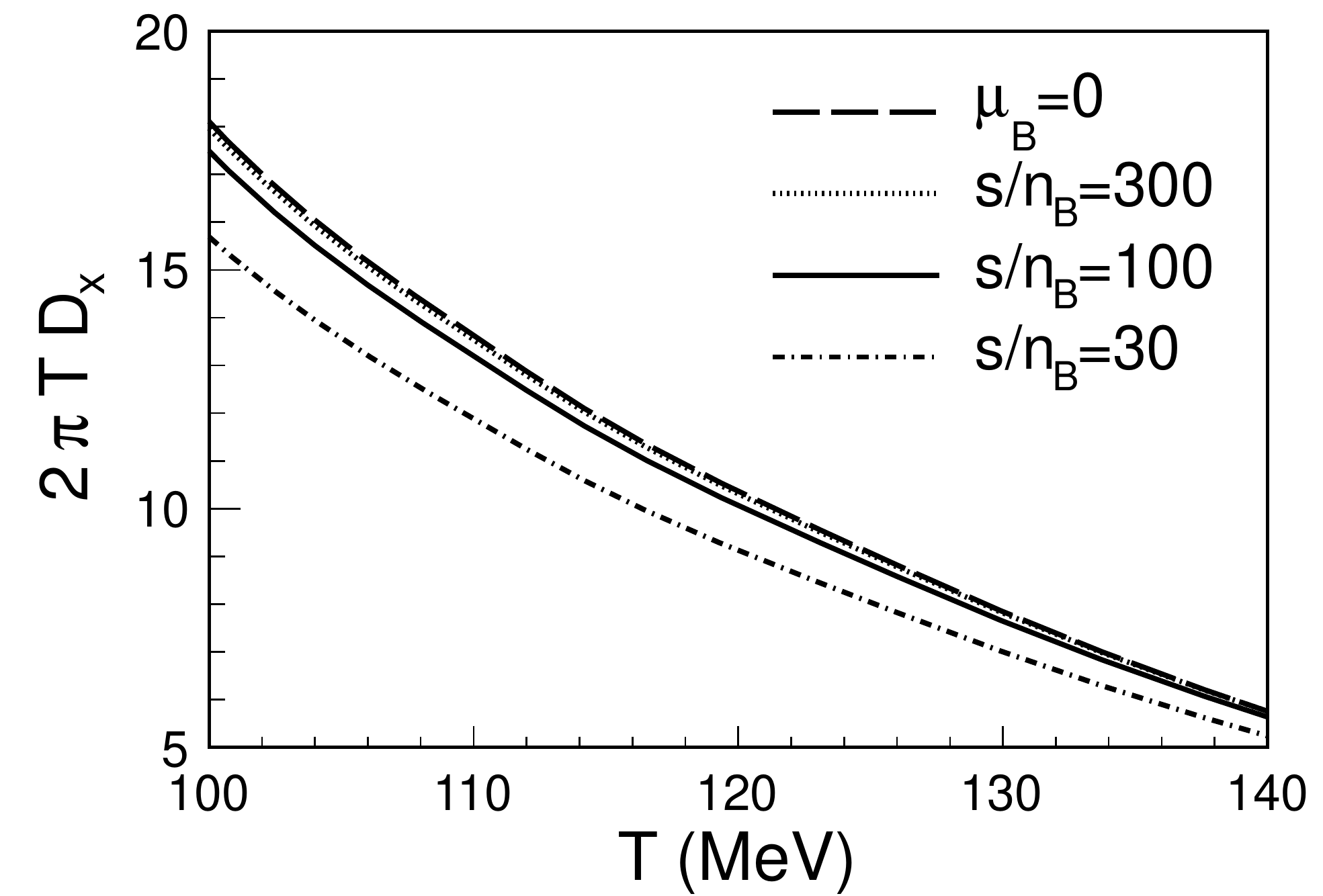}
\caption{\label{fig:tau_R}
Spatial diffusion coefficient multiplied by the thermal wavelength  ($2 \pi T$)  for the different isentropic curves shown in Fig.~\ref{fig:isen}.
}
\end{figure}

In the upper panel of Fig.~\ref{fig:tau_R} we show the relaxation time $\tau_{R}=1/F$. Because $\tau_R$ is 
much larger than the lifetime of the system~\cite{Adams:2004yc,Aamodt:2011mr}, the bottom can hardly relax during the
fireball expansion. In other words, the collisions with other particles are not enough to appreciably reduce the average
momentum. The three curves are quite similar in the whole range of temperatures. In fact, we have added the limiting case of $\mu_B=0$ that corresponds
to $s/n_{B} \rightarrow \infty$ and check that it is almost indistinguishable from the curve of
$s/n_{B}=100$. For collisions with larger baryonic density (lower entropy per baryon) the relaxation time is smaller, because the
heavy meson scatters more, but not enough to represent a efficient mechanism of relaxation.

One can also compare these results to those for the charm case in Ref.~\cite{Tolos:2013kva}.
The drag coefficient naively scales with the inverse mass of the heavy meson
\be \label{eq:nonre} F \sim P \sigma \sqrt{\frac{m_l}{T}} \frac{1}{m_B} \ , \ee
where $P$ is the pressure of the bath, $\sigma$ is the total cross section and $m_l$ is the mass of the bath's particles.
Assuming comparable interactions (i.e. similar cross sections), we expect $\tau_R( {\rm bottom})/\tau_R ({\rm charm}) \sim m_B/m_D \simeq 2.8$.
Comparing with the curve of $s/n_{B}=30$ in Ref.~\cite{Tolos:2013kva}, one can check that this is
indeed satisfied (the breaking of this scaling can be accounted by differences in the cross sections).
As a representative value for the bottom case we can quote a relaxation time of $\tau_R ({\rm bottom}) = 67.9$ fm for $T=140$ MeV at
$s/n_{B}=30$. This value has to be compared with $\tau_R ({\rm charm})= 28.3$ fm at the same temperature and entropy per baryon \cite{Tolos:2013kva}.

Finally, in the lower panel of Fig.~\ref{fig:tau_R} we plot the spatial diffusion coefficient
$D_x$. This coefficient is multiplied by the thermal wavelength ($2 \pi T$) to form an adimensional number, like the Reynolds or the Knudsen numbers.
The results are again quite independent of the entropy per baryon as long as the collision energy
is high enough. In conclusion, our results can be taken as prediction for the hadronic medium created at 
high energy collisions (like those at the RHIC or the LHC) independently of the precise value
of the entropy per baryon of the trajectory.

\section{Conclusions \label{sec:conclusions}}

We have studied the interaction and propagation of $\bar B$ mesons in hadronic matter made of light mesons, $N$ and $\Delta$,  by means of a unitarized approach based on
effective models that are compatible with chiral and heavy quark symmetries. 

We have examined the $\bar B$ scattering with mesons and baryons by analyzing the dynamically generated states in the meson and baryon sectors where the $\bar B$ meson and its
HQSS partner, the $\bar B^*$ meson, are present. In the open bottom meson sector there is a strong resemblance between the $0^+$ and $1^+$ spectrum due to HQSS. Among others, we have
found two non-strange resonances that form a HQSS doublet, the $B_0(5530)$ and $B_1(5579)$ states, which turn out to be the bottom counterparts of the experimental $D_0(2400)$ and $D_1(2430)$, respectively. These
resonances have not been experimentally observed yet. Moreover, another doublet with strangeness, $B_{s0}^*(5748)$ and $B_{s1}^*(5799)$, can be identified as the bottom analogues of the $D_{s0}^* (2317)$ and the $D_{s1} (2460)$. For baryons we have also determined several
$J=1/2$ and $J=3/2$ states in the $\Lambda_b$ and $\Sigma_b$ sectors which form HQSS doublets. This is the case of the $\Lambda_b(5910)$ and $\Lambda^*_b(5921)$, which can be identified with the states
observed by the LHCb collaboration \cite{Aaij:2012da}. Furthermore, we have associated one of our states, the $J=3/2$ $\Sigma_b^*(5904)$ to be the bottom counterpart of the strange $\Sigma^*(1670)$ and
charmed $\Sigma_c^*(2549)$ resonances, though not experimentally detected yet but a clear case for discovery.

Next we have analyzed different transport coefficients that describe the propagation of $\bar B$ mesons in hadronic matter. We have shown the drag and diffusion coefficients for vanishing
baryochemical potential, which serve as inputs for the numerical propagation of the heavy meson in matter produced at high-energy colliders like RHIC or LHC. At $\mu_B=0$ the main contribution to the drag
and diffusion coefficients comes from the interaction of $\bar B$ mesons with pions as the thermal bath is mainly populated by this species. Alternatively, we have also presented other
quantities that possess a more physical insight, such as  the relaxation time $\tau_R$ and the spatial diffusion coefficient $D_x$ for isentropic trajectories within the QCD phase diagram. These trajectories
range from the region explored by the RHIC and LHC experiments up to FAIR at its top energy. We have checked that  the naive scaling of the relaxation time with the inverse mass of the heavy meson
is fulfilled. Moreover, although the relaxation time is smaller with larger baryonic density, the $\bar B$ meson can hardly relax to the equilibrium. Indeed, our results can be taken as
predictions for the hadronic medium created at high energy collisions (like those at the RHIC or the LHC) independently of the precise value of the entropy per baryon of the trajectory as long as the collision energy is high enough.

\begin{acknowledgments}
  We would like to thank L. Abreu for providing us with the coefficients of Table~\ref{tab:isoscoeff}.
This work has been funded by Grants No. FPA2010-16963 (Ministerio de Ciencia e Innovaci\'on) and No. FP7-PEOPLE-2011-CIG under Contract No. PCIG09-GA-2011-291679. We acknowledge the support of the European Community-Research Infrastructure
Integrating Activity Study of Strongly Interacting Matter (acronym HadronPhysics3, Grant Agreement n. 283286) under the Seventh Framework Programme of EU. L.T. acknowledges support from the Ram\'on y Cajal
Research Programme (Ministerio de Ciencia e Innovaci\'on). J.M.T.-R. is also supported by the Programme TOGETHER from R\'egion Pays de la Loire and the European I3-Hadron Physics programme.
\end{acknowledgments}

\end{document}